\documentclass[prd, preprint, longbibliography, 12pt]{revtex4-1}
\usepackage{amsmath}
\usepackage{amssymb}
\usepackage{setspace}
\usepackage{graphicx}
\usepackage{natbib}
\usepackage{float}
\usepackage{tensor}
\usepackage[utf8]{inputenc}
\usepackage{amsfonts}
\usepackage{braket}
\usepackage{esint}
\usepackage{breqn}
\usepackage{IEEEtrantools}

\begin{document}
 
 %

\begin{center}

\large{\bf Quantum gravity effects in the infra-red: a theoretical derivation of the low energy fine structure constant and mass ratios of elementary particles}


\vskip 0.2 in

{\large{\bf Tejinder P.  Singh }}

\medskip

{\it Tata Institute of Fundamental Research,}
{\it Homi Bhabha Road, Mumbai 400005, India}\\
\bigskip
  { \tt e-mail: tpsingh@tifr.res.in}
\bigskip

\end{center}

\vskip 0.1 in

\centerline{\bf Accepted for publication in  The European Physical Journal Plus}

\vskip 0.2 in

\bigskip

\centerline{\bf Abstract}
\noindent
We have recently proposed a pre-quantum, pre-space-time theory as a matrix-valued Lagrangian dynamics on an octonionic space-time. This theory offers the prospect of unifying  internal symmetries of the standard model with pre-gravitation. We explain why such a quantum gravitational dynamics is in principle essential even at energies much smaller than Planck scale.  The dynamics  can also predict the values of free parameters of the low energy standard model: these parameters arising in the Lagrangian are related to the algebra of the octonions which define the underlying non-commutative space-time on which the dynamical degrees of freedom evolve. These free parameters are related to the exceptional Jordan algebra $J_3(8)$ which describes the three fermion generations. We use the  octonionic representation of  fermions to compute the eigenvalues of the characteristic equation of  this algebra, and compare the resulting eigenvalues with known mass ratios for quarks and leptons. We show that the ratios of the eigenvalues correctly reproduce the [square root of the] known  mass ratios.  In conjunction with the trace dynamics Lagrangian, these eigenvalues also yield  a theoretical derivation of the low energy  fine structure constant.

\newpage

\tableofcontents

\section{Unification of gravitation with the standard model is needed at all energy scales, not just at the Planck energy scale}
Several theoretical and experimental physicists are interested in the following question: if a massive object were to be created in a quantum state which is a superposition of two position states, what is the gravitational field that it produces? One can expect, on physical grounds, that such a gravitational field will be non-classical, i.e. not describable by Newton's laws of gravitation. In that sense, the field is quantum gravitational, and is an example of low-energy quantum gravity, i.e. quantum gravity in the infra-red. Furthermore, the system is non-relativistic, so we can call it non-relativistic low-energy quantum gravity. Such quantum gravity is of course different from Planck energy scale quantum gravity, which is at very high energies and also relativistic. However, we can prescribe unifying criteria which unify both these limits into a common description. The gravitational field produced by a quantum system having an action $S$ of the order $\hbar$ is quantum gravitational in nature and will exhibit a quantum superposition of spacetime geometries. Because, only when $S\gg \hbar$ does the quantum system achieve its classical limit, and in this limit the associated gravitational field also becomes classical. Next, if the time scale $T$ associated with this system is of the order Planck time $t_{Pl}$, then the associated energy scale $\hbar /T$ is of the order Planck energy. However, if $T\gg t_{Pl}$, as is true in present-day laboratory experiments, the associated energy scale is much less than Planck energy. Nonetheless, this is a quantum gravitational system, because the source is quantum in nature. Such low energy quantum gravity will be non-relativistic weak quantum gravity if the source is moving slowly, as in the aforesaid superposition of a mass in two places. However if the quantum source is relativistic, we have relativistic weak quantum gravity, an example of which is a relativistic low energy electron obeying the quantum field theoretic laws of the standard model. 

Quantum systems that are described by the standard model are a source of relativistic weak quantum gravity. However, the gravitational field they produce cannot be described by perturbative quantum field theory around a Minkowski spacetime background: because a chosen quantum state is in general a superposition of many position states, and there will be a classical gravitational field corresponding to each of those positions: which position to choose as the preferred one, around which to define the field? Nor can the said field be a superposition of such classical fields, because  then there is no background spacetime left around which to carry out a quantum field theoretic analysis. Another way to put it is to say that when all sources are quantum, the coordinate geometry of spacetime cannot be described by real numbers, and the point structure is lost. This of course is the notorious tension between quantum superposition and well-defined classical spacetime geometry, and is a consequence of the Einstein hole argument \cite{Singh:2012,  singhessay}. To resolve this conflict we must invoke a 
non-commutative coordinate geometry, even at low energies, and  proceed as follows.

Just as the Dirac equation is the square-root of the Klein-Gordon equation, a spinor spacetime is the square-root of Minkowski spacetime. 
If we are not interested in the gravitational field of an electron, it is perfectly fine to work with the Dirac equation written on a Minkowski spacetime.
However, if we want to know the gravitational field produced by the electron, we must first define the electron states on a spinor spacetime.
Octonions define a spinor spacetime, which has eight non-commuting dimensions. Its square is a ten-dimensional  Minkowski spacetime, by virtue of the homomorphism $SL(2, \mathbb{O}) \sim SO(1,9)$.
Using Clifford algebras, the spinorial states for fermions can be defined on 8D octonionic spacetime. This is the construction for three fermion generations which will be described subsequently in the paper. The exceptional Lie group $E_6$ is the symmetry group of the Dirac equation in 10D spacetime \cite{DrMa, mdw} and it is also the automorphism group of the complexified exceptional Jordan algebra. The group $F_4$ is the automorphism group of the exceptional Jordan algebra $J_3(8)$ of $3\times 3$ Hermitean matrices with octonionic entries. The eigenvalues of this algebra therefore are a solution to the eigenvalue problem for the Dirac equation in 10D when defined on a spinor equivalent of Minkowski spacetime. 

The symmetries of the octonionic space restrict what properties the fermions can have. Charge and (square-root) mass are both defined as eigenvalues of $U(1)$ symmetry operators of the 8D space, and take discrete values consistent with what is observed experimentally in the standard model. The allowed properties show that the only fermions possible are left-handed and right-handed quarks and leptons of the three generations. This includes three right handed sterile neutrinos, one per generation.
The gravitational effect of an electron is equivalent to curving of this 8D octonionic space-time, and is described by the dynamical equations of a generalised  trace dynamics \cite{Singhreview}. 
The description of the standard model using the laws of QFT on Minkowski spacetime, while extremely successful, is an approximate description. It does not tell us why the standard model is what it is, and why the dimensionless constants take those particular values which we see in experiments.
On the other hand, when we describe the dynamics of elementary particles using trace dynamics on a spinor spacetime, the symmetries of the standard model and its dimensionless constants are determined by the algebraic properties of the 8D octonionic spacetime. There is no freedom. The symmetry group of the theory is $E_6$ and permits the extension of the standard model to include the right-handed sector, which is the (pre-)gravitational counterpart of the standard model.

This description in terms of a spinor spacetime is available and essential at all energy scales, low as well as high. That is the reason why the low energy fine structure constant and mass ratios get determined in this theory. The Jordan eigenvalues give the expansion of the left-handed charge eigenstates in terms of the right-handed square-root mass eigenstates, and vice versa. This is the reason these eigenvalues yield a theoretical determination of the mass-ratios. By a pre-spacetime pre-quantum theory we do not just mean a pre-theory at Planck scale energies. This pre-theory is also essential at low energies for us to understand the standard model at low energies. We call this relativistic weak quantum gravity coupled to the standard model.
QFT on 4D Minkowski spacetime can be recovered from trace dynamics on a spinor spacetime under appropriate conditions \cite{Singhreview}.

\section{Division algebras, Clifford algebras, and the standard model}
We have recently proposed a pre-quantum, pre-space-time theory, which is a matrix-valued Lagrangian dynamics, written on an octonionic space-time \cite{Singhreview}. This theory generalises Adler's theory of trace dynamics \cite{Adler:94, AdlerMillard:1996, Adler:04}, which is a pre-quantum theory on a four-dimensional  Minkowski space-time \cite{Singh2020DA, GRFEssay2021}. It is a Lagrangian dynamics for Yang-Mills fields, fermions, and gravity. The algebra automorphisms of the octonions, which form the smallest exceptional Lie group $G_2$, play the role of unifying general coordinate transformations (i.e. space-time diffeomorphisms) with internal gauge transformations. We wrote down the Lagrangian for one generation of standard model fermions and gauge bosons, in this pre-theory. A Clifford algebra $Cl(6, C)$ constructed from the octonion algebra is used to make spinors [`minimum left ideals' of $Cl(6, C)]$ which represent the eight fermions of one generation, and their anti-particles, and their electro-color symmetry. Another $Cl(6, C)$ made from the octonions describes the action of the Lorentz-weak symmetry on these octonions. These aspects of one-generation of fermions are confirmed by the Lagrangian dynamics constructed in the pre-theory. Our results are in agreement with the earlier work of Furey \cite{f1, f2, f3} and Stoica \cite{Stoica} for the Clifford algebra $Cl(6, C)$ based description of one generation of standard model fermions. In our work, quantum field theory of the standard model emerges from the pre-theory, at energies much lower than the Planck scale. The Appendix in Section V below summarises the theoretical background of the present article, as developed in our earlier papers \cite{maithresh2019, Singhspin, MPSingh, Singh2020DA}. The present paper should ideally be read as a continuation of \cite{Singh2020DA}. We explain how the octonionic space-time, on which the fermions reside, fixes the dimensionless free parameters of the standard model [which appear in the octonionic Lagrangian]  as a consequence of the properties of the algebra of the octonions \cite{Vatsalya1}, this being the exceptional Jordan algebra $J_3(8)$.

The possible connection between division algebras, exceptional Lie groups, and the standard model has been a subject of interest for many researchers in the last few decades \cite{Dixon, Gursey, f1, f2, f3, Chisholm, Trayling, Dubois_Violette_2016, Todorov:2019hlc, Dubois-Violette:2018wgs, Todorov:2018yvi, Todorov:2020zae, ablamoowicz, baez2001octonions, Baez_2011,  baez2009algebra, f1, f2, f3, Perelman,  Gillard2019, Stoica, Yokota, Dray1, Dray2, lisi2007exceptionally, Ramond1976}.  Our own interest in this connection stems from the following observation \cite{Singh2020DA}. In the pre-geometric, pre-quantum theory of generalised trace dynamics, the definition of spin requires 4D space-time to be generalised to an 8D non-commutative space. In this case, an octonionic space is a possible, natural, choice for further investigation. We found that the additional four directions can serve as `internal' directions and open a path towards a possible unification of the Lorentz symmetry with the standard model, with gravitation arising only as an emergent phenomenon. Instead of the Lorentz transformations and internal gauge transformations, the symmetries of the octonionic space are now described by the automorphisms of the octonion algebra.  Remarkably enough, the symmetry  groups of this algebra, namely the exceptional Lie groups, naturally have in them the desired symmetries [and {\it only those} symmetries, or higher ones built from them] of the standard model, including Lorentz symmetry, without the need for any fine tuning or adjustments. Thus the group of automorphisms of the octonions is $G_2$,  the smallest of the five exceptional Lie groups $G_2, F_4, E_6, E_7, E_8$. The group $G_2$ has two intersecting maximal sub-groups \cite{tkey}, $SU(3) $ and $SO(4)$, which between them account for the fourteen generators of $G_2$, and can possibly serve as the symmetry group for one generation of standard model fermions.  The complexified Clifford algebra $Cl(6, C)$ plays a very important role in establishing this connection. In particular, motivated by a map between the complexified octonion algebra and $Cl(6, C)$, electric charge is defined as one-third the eigenvalue of a $U(1)$ number operator, which is identified with $U(1)_{em}$ \cite{f1, f3}.

Describing the symmetries $SU(3) \times U(1)$ and $SU(2)_L \times SU(2)_R$ of the standard model [with Lorentz symmetry now included, through extension of $SU(2)_R$ to $SL(2,C)$ \cite{Vatsalya1}]  requires two copies of the Clifford algebra $Cl(6, C)$ whereas the octonion algebra yields only one such independent copy. It turns out that if boundary terms are not dropped from the Lagrangian of our theory, the Lagrangian describes three fermion generations [\cite{Singh2020DA} and Section III below in the present paper], with the symmetry group now raised to $E_6$, and to $F_4$ for determining parameter values. This admits three intersecting copies of $G_2$, with the $SU(2) \times SU(2)$ in the intersection, and a Clifford algebra construction based on the three copies of the octonion algebra is now possible \cite{singhessay, Gillard}.  Attention thus shifts to investigating the connection between $F_4$ and the three generations of the standard model.

$F_4$ is also the group of automorphisms of the exceptional Jordan algebra \cite{Albert1933, Jordan, Todorov:2018yvi}. The elements of the algebra are 3x3 Hermitean matrices with octonionic entries. This algebra admits an important cubic characteristic equation with real eigenvalues. Now we know that the three fermion generations differ from each other only in the mass of the corresponding fermion, whereas the electric charge remains unchanged across the generations. This motivates us to ask: if the eigenvalues of the $U(1)$ number operator constructed from the octonion algebra represent electric charge, what is represented by the eigenvalues of the exceptional Jordan algebra? Could these eigenvalues bear a relation with mass ratios of quarks and leptons? This is the question investigated in the present paper and answered in the affirmative. Using the very same octonion algebra which was used to construct a state basis for standard model fermions, we calculate these eigenvalues. Remarkably, the eigenvalues are very simple to express, and bear a simple relation with electric charge. We describe how they  relate to mass ratios. In particular we find that the ratios of the eigenvalues match with the square root of the mass ratios of charged fermions. [These eigenvalues are invariant under algebra automorphisms, the automorphism group being $F_4$, and the automorphisms of one chosen coordinate representation of the fermions, as below, give other equivalent coordinate representations for the same set of fermions. Octonions serve as coordinate systems on the eight dimensional octonionic space-time manifold on which the elementary fermions live. The Appendix at the end of this paper reviews this 8D space-time picture]. 

Thus we are asking that when the octonions representing the three fermion generations are used as the off-diagonal entries in the 3x3 Jordan matrices, and the diagonal entries are the electric charges, what is the physical interpretation of the eigenvalues of the characteristic equation of $J_3(\mathbb O)$? These eigenvalues are made from the invariants of the algebra, and hence are themselves invariants. So they are likely to carry significant information about the standard model. This is what we explore in the present paper, and we argue that these eigenvalues inform us about mass-ratios of elementary particles, and about the coupling constants of the standard model.

Subsequently in the paper we propose a diagrammatic representation, based on octonions and $F_4$, of the fourteen gauge bosons, and the (8x2)x3 = 48 fermions of three generations of standard model, along with the  Higgs. We attempt to explain why there are not three generations of bosons, and re-express our Lagrangian in a form which explicitly reflects this fact. We also argue as to how this Lagrangian might directly lead to the characteristic equation of the exceptional Jordan algebra, and reveal why the eigenvalues might be related to mass. Furthermore, we identify the standard model coupling constants in our Lagrangian, and by relating them to the eigenvalues of $J_3(\mathbb O)$ we provide a theoretical derivation of the asymptotic fine structure constant value 1/137.xxx We also note that the determination of these eigenvalues is equivalent to solving the Dirac equation for three fermion generations in 10D Minkowski spacetime, the symmetry group being $E_6$. Also, what we call the octonionic spacetime can be regarded as an octonion valued (instead of complex number valued) twistor space, reaffirming Penrose's proposal that in quantum gravity, spacetime is a spinor spacetime.

It is known that since $F_4$ does not have complex representations, it cannot give a representation of the fermion states. It has hence been suggested that the correct representation could come from the next exceptional Lie group, $E_6$, which is the automorphism group of the complexified exceptional Jordan algebra. This aspect is currently being investigated by several researchers, including the present author. However, the standard model free parameters certainly cannot come from the characteristic equation related to $E_6$, because the roots of this equation are not real numbers in general. It is clear that the parameters must then come from the roots of the characteristic equation of $F_4$, which in a sense is the 
self-adjoint counterpart  of the equation for $E_6$. It is in this spirit that the present investigation is carried out, and the results we find suggest that the present approach is indeed the correct one, as regards determining the model parameters. One must investigate $E_6$ for representations, but $F_4$ for the parameter values. The exceptional Lie group $E_6$ is the symmetry group of the Dirac equation in 10D spacetime \cite{DrMa}.  The eigenvalues of the algebra $J_3(8)$ therefore are a solution to the eigenvalue problem for the Dirac equation in 10D when defined on a spinor equivalent of Minkowski spacetime. 

The plan of the paper is as follows. In the next section we recall the exceptional Jordan algebra, construct the octonionic representation of the three fermion generations, calculate the roots of the characteristic equation, and make some comments on mass-ratios and the roots. In Section IV we construct the trace dynamics Lagrangian for three generations, along with the bosons, and we give a theoretical derivation of the asymptotic fine structure constant from first principles. In Section V we explain how the  Jordan eigenvalues in fact act as a definition of mass, quantised in units of Planck mass. We then show that mass ratios of charged fermions are obtained from these eigenvalues. In the Appendix in Section VI we recall the motivation in earlier work, for developing this pre-theory, and we also include a few new insights. In particular we report on a 4D quaternionic version of the pre-theory, which describes the Lorentz-weak interaction of the leptons, based on an extension of the Lorentz algebra by $SU(2)_L$. In order to include quarks and the strong interaction, this 4D quaternionic pre-theory is extended to eight octonionic dimensions.

\section{Three fermion generations, and physical eigenvalues from the characteristic equation of the Exceptional Jordan Algebra}
The exceptional Jordan algebra  [EJA] $J_3({\mathbb O})$ is the algebra of 3x3 Hermitean matrices with octonionic entries \cite{Albert1933, Dray1, Dray2, Todorov:2020zae}
\begin{equation}
X(\xi, x)=
\begin{bmatrix}
\xi_1 &  x_3 & x_2^*\\
x_3^* & \xi_2 & x_1\\
x_2 & x_1^* & \xi_3
\end{bmatrix}
\end{equation}
It satisfies the characteristic equation \cite{Todorov:2020zae, Dray1, Dray2}
\begin{equation}
X^3 - Tr (X) X^2 + S(X) X - Det(X) = 0; \qquad Tr(X) = \xi_1 + \xi_2 + \xi_3 \qquad 
\end{equation}
which is also satisfied by the eigenvalues $\lambda$ of this matrix
\begin{equation}
\lambda^3 - Tr(X) \lambda^2 + S(X) \lambda - Det(X)=0
\end{equation}
Here the determinant is
\begin{equation}
Det (X) = \xi_1 \xi_2 \xi_3 + 2 Re (x_1 x_2 x_3) - \sum_1^3 \xi_i x_i x_i^*
\end{equation} 
and $S(X)$ is given by
\begin{equation}
S(X) = \xi_1 \xi_2 - x_3 x_3^* + \xi_2 \xi_3 - x_1 x_1^* + \xi_1 \xi_3 - x_2^* x_2
\end{equation}
The diagonal entries are real numbers and the off-diagonal entries are (real-valued) octonions. A star denotes an octonionic conjugate.
The automorphism group of this algebra is the exceptional Lie group $F_4$. Because the Jordan matrix is Hermitean, it has real eigenvalues which can be obtained by solving the above-given eigenvalue equation.

In the present article we suggest that these eigenvalues carry information about mass ratios of quarks and leptons of the standard model, provided we suitably employ the octonionic entries and the diagonal real elements to describe quarks and leptons of the standard model. Building on earlier work \cite{f1, f2, Stoica} we recently showed that the complexified Clifford algebra $Cl(6, C)$ made from the octonions acting on themselves can be used to obtain an explicit octonionic representation for a single generation of eight quarks and leptons, and their anti-particles. 
In a specific basis, using the neutrino as the idempotent $V$, this representation is as follows \cite{f1, Singh2020DA}. The $\alpha$ are fermionic ladder operators of $Cl(6, C)$ (please see Eqn. (34) of \cite{Singh2020DA}). 
\begin{equation}
\begin{split}
V = \frac{i}{2} e_7 \qquad [V_\nu  \  {\rm Neutrino}]\\
\alpha_1^\dagger V = \frac{1}{2} ( e_5 + ie_4)\times V = \frac{1}{4} ( e_5 + ie_4) \qquad [\rm V_{ad1}\  Anti-down\ quark] \\
\alpha_2 ^\dagger V = \frac{1}{2} ( e_3 + ie_1)\times V = \frac{1}{4} ( e_3 + ie_1) \qquad [\rm V_{ad2}\ Anti-down\ quark] \\
\alpha_3^\dagger V = \frac{1}{2} ( e_6 + ie_2)\times V = \frac{1}{4} ( e_6 + ie_2) \qquad [\rm V_{ad3} \ Anti-down\ quark] \\
\alpha_3^\dagger \alpha_2^\dagger V = \frac{1}{4} ( e_4+ ie_5) \qquad [\rm V_{u1}\ Up\ quark] \\
\alpha_1^\dagger \alpha_3^\dagger V =  = \frac{1}{4} ( e_1 + ie_3) \qquad [\rm V_{u2}\ Up\ quark] \\
\alpha_2^\dagger \alpha_1^\dagger V =  \frac{1}{4} ( e_2 + ie_6) \qquad [\rm V_{u3}\ Up\ quark] \\
\alpha_3^\dagger \alpha_2^\dagger \alpha_1^\dagger V = -\frac{1}{4}(i+e_7) \qquad [{\rm V_{e+}\ Positron}]
\end{split}
\label{expr}
\end{equation}
The anti-particles are obtained from the above representation by complex conjugation \cite{f1}. 

Note: Eqn. (33) of \cite{Singh2020DA} for the idempotent has an incorrectly written expression on the right hand side. Instead of $ie_7/2$ as written there, the correct expression is $(1+ie_7)/2$ \cite{Mondal}. Hence the idempotent $V$ in that paper should be $(1+ie_7)/2$, not $ie_7/2$. It has now been found however, that identification of the neutrino with the idempotent $V=(1+ie_7)/2$ does not give the desired values for mass-ratios and coupling constants reported in the present paper \cite{Mondal}. We hence propose the Majorana particle interpretation for the neutrino, and identify the neutrino with $(V-V_{cc})/2$ where $V_{cc}$ is the complex conjugate of $V$. Hence the neutrino is $[(1+ie_7)-(1-ie_7)]/4 = ie_7/2$, so that the octonionic representation of the neutrino remains the same as shown in \cite{Singh2020DA} and is the one used in the present paper. Our results here seem to suggest that the neutrino is a Majorana particle, and not a Dirac particle.

Note: In Eqn. (34) of \cite{Singh2020DA} the denominator in the expression for the positron should be 4, not 8. The correct expression for the positron is shown above in Eqn. (\ref{expr}). 

In the context of the projective geometry of the octonionic projective plane ${\mathbb OP^2}$ it has been shown by Baez \cite{baez2001octonions} that upto automorphisms, projections in EJA take one of the following four forms, having the respective invariant trace $0, 1, 2, 3$.
\begin{equation}
p_0 =
\begin{bmatrix}
0 &  0 & 0\\
0 & 0 & 0\\
0 & 0 & 0
\end{bmatrix}
\end{equation}
\begin{equation}
p_1 =
\begin{bmatrix}
1 &  0 & 0\\
0 & 0 & 0\\
0 & 0 & 0
\end{bmatrix}
\end{equation}
\begin{equation}
p_2 =
\begin{bmatrix}
1 &  0 & 0\\
0 & 1 & 0\\
0 & 0 & 0
\end{bmatrix}
\end{equation}
\begin{equation}
p_3=
\begin{bmatrix}
1 &  0 & 0\\
0 & 1 & 0\\
0 & 0 & 1
\end{bmatrix}
\end{equation}
Since it has earlier been shown by Furey \cite{f1} that electric charge is defined in the division algebra framework as one-third of the eigenvalue of a $U(1)$ number operator made from the generators of the $SU(3)$ in $G_2$, we propose to identify the trace of the Jordan matrix with the sum of the charges of the three identically charged fermions across the three generations. Thus the trace zero Jordan matrix will have diagonal entries zero, and will represent the (neutrino, muon neutrino, tau-neutrino). The trace one Jordan matrix will have diagonal entries $(1/3, 1/3, 1/3)$ and will represent the (anti-down quark, anti-strange quark, anti-bottom quark). [Color is not relevant for determination of mass eigenvalues, and hence effectively we have four fermions per generation: two leptons and two quarks, after suppressing color]. The trace two Jordan matrix will have entries $(2/3, 2/3, 2/3)$ and will represent the (up quark, charm, top). Lastly, the trace three Jordan matrix will have entries (1, 1, 1) and will represent (positron, anti-muon, anti-tau-lepton). 

We have thus identified the diagonal real entries of the four Jordan matrices whose eigenvalues we seek. We must next specify the octonionic entries in each of the four Jordan matrices. Note however that the above representation of the fermions of one generation is using complex octonions, whereas the entries in the Jordan matrices are real octonions. So we devise the following scheme for a one-to-one map from the complex octonion to a real octonion. Since we are ignoring color, we pick one out of the three up quarks, say $(e_4 + i e_5)$, and one of three anti-down quarks, say $(e_5 + i e_4)$. Since the representation for the electron and the neutrino use $e_7$ and a complex number, it follows that the four octonions we have picked form the quaternionic triplet $(e_4, e_5, e_7)$ [we use the Fano plane convention shown Fig. 1 below]. Hence the four said octonions are in fact complex quaternions, thus belonging to the general form
\begin{equation}
(a_0 + i a_1) + (a_2 + i a_3) e_4 + (a_4 + i a_5) e_5 + (a_6+ia_7) e_7
\end{equation} 
where the eight $a$-s are real numbers. By definition, we map this complex quaternion to the following real octonion:
\begin{equation}
a_0 + a_1 e_1 + a_5 e_2 + a_3 e_3 + a_2 e_4 + a_4 e_5 + a_7 e_6 + a_6 e_7
\end{equation}
Note that the four real coefficients in the original complex quaternion have been kept in place, and their four imaginary counterparts have been moved to the octonion directions $(e_1, e_2, e_3, e_6)$ now as real numbers. Clearly, the map is reversible, given the real octonion we can construct the equivalent complex quaternion representing the fermion.
 \begin{figure}[!htb]
        \center{\includegraphics[width=\textwidth]
        {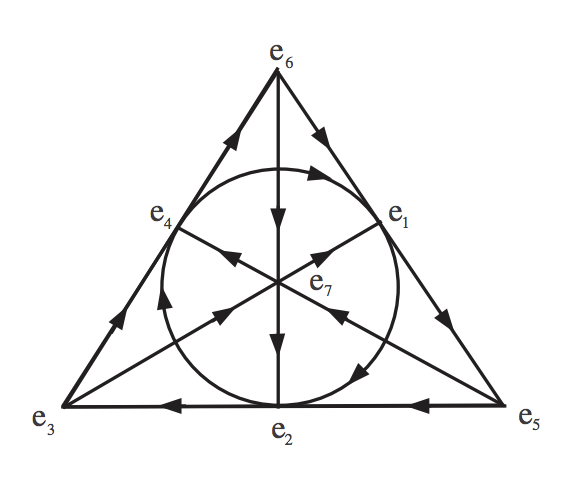}}
        \caption{\label{fig:my-label} The Fano plane.}
      \end{figure}
We can now use this map and construct the following four real octonions for the neutrino, anti-down quark, up quark and the positron, respectively, after comparing with their complex octonion representation above.
\begin{equation} 
V_{\nu} = \frac{i}{2} e_7 \longrightarrow \frac{1}{2} e_6
\end{equation}
\begin{equation}
V_{ad} = \frac{1}{4} e_5 + \frac{i}{4} e_4 \longrightarrow \frac{1}{4} e_5 + \frac{1}{4} e_3
\end{equation} 
\begin{equation}
V_{u} = \frac{1}{4} e_4 + \frac  {i}{4} e_5 \longrightarrow \frac{1}{4} e_4 + \frac{1}{4} e_2
\end{equation}
\begin{equation}
V_{e^{+}} = - \frac{i}{4} - \frac{1}{4} e_7 \longrightarrow - \frac{1}{4} e_1 -\frac{1}{4} e_7
\end{equation}
These four  real octonions will go, one each, in the four different Jordan matrices whose eigenvalues we wish to calculate. Next, we need the real octonionic representations for the four fermions [color suppressed] in the second generation and the four in the third generation. We propose to build these as follows, from the real octonion representations made just above for the first generation. Since $F_4$ has the inclusion $SU(3) \times SU(3)$, one $SU3)$ being for color and the other for generation, we propose to obtain the second generation by a 
$2\pi /3$ rotation on the first generation, and the third generation by a $2\pi / 3$ rotation on the second generation. By this we mean the following construction, for the four respective Jordan matrices, as below. It is justified as follows: One of the two $SU(3)$ is color $SU(3)_c$ and has already been used up to write down the three different color states of each quark, with one pair of imaginary octonion directions fixed for a given color.  The other $SU(3)$ is for generations. It is then evident from symmetry considerations that the corresponding higher generation quark of a given color can be obtained by $2\pi/3$ rotation on the first generation quark, while keeping the selected pair of octonionic directions fixed.  

Up quark / Charm / Top: The up quark is $(e_4 /4 + e_2 /4)$ We think of this as a `plane' and rotate this octonion by $2\pi/3$ by left multiplying it by $e^{2\pi e_4/3}= -1/2 + \sqrt{3} e_4 /2$. 
This will be the charm quark $V_c$. Then we left multiply the charm quark by $e^{2\pi e_4/3}$ to get the top quark $V_t$. Hence we have,
\begin{equation}
V_c = (-1/2 + \sqrt{3} e_4/2) \times V_u = (-1/2 + \sqrt{3} e_4 /2) \times \left(\frac{1}{4} e_4 + \frac{1}{4} e_2\right) = -\frac{1}{8} e_4 - \frac{1}{8} e_2 - \frac{\sqrt{3}}{8} - \frac{\sqrt{3}}{8} e_1
\end{equation}
 \begin{figure}[!htb]
        \center{\includegraphics[width=\textwidth]
        {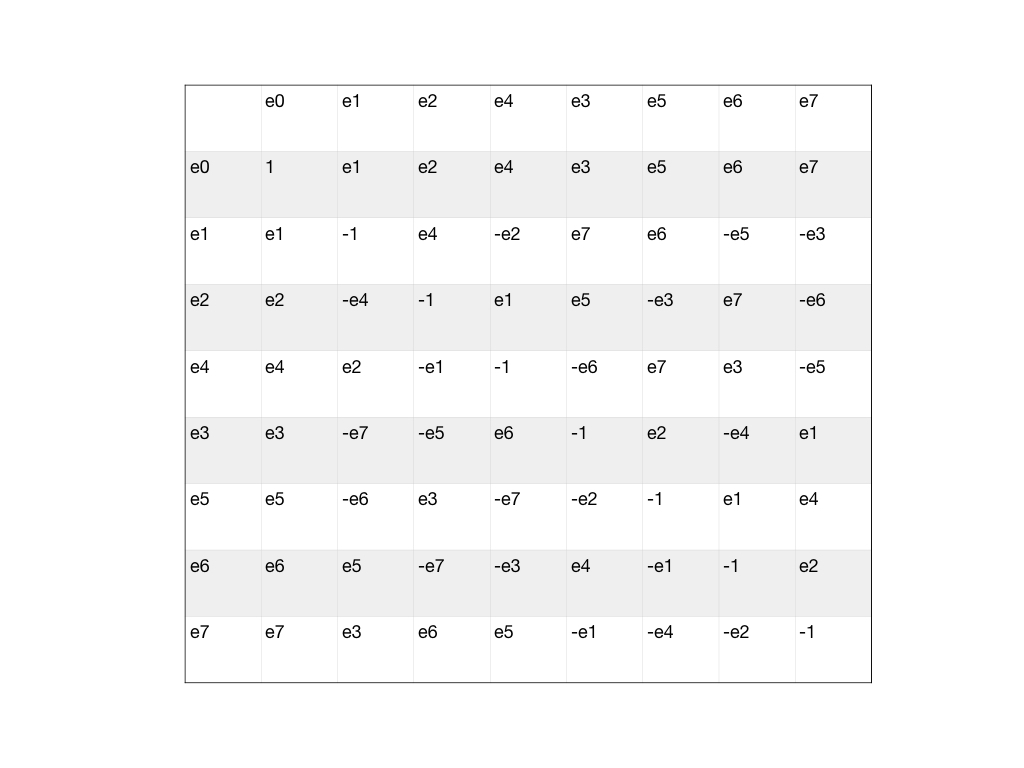}}
        \caption{\label{fig:my-label} The multiplication table for two octonions. Elements in the first column on the left, left multiply elements in the top row.}
      \end{figure}
We have used the conventional multiplication rules for the octonions, which are reproduced  in Fig. 2, for ready reference. Similarly, we can construct the top quark by a $2\pi/3$ rotation on the charm:
\begin{equation}
\begin{split}
V_t & = (-1/2 + \sqrt{3} e_4 /2) \times V_c = (-1/2 + \sqrt{3} e_4 /2) \times \left(-\frac{1}{8} e_4 - \frac{1}{8} e_2 - \frac{\sqrt{3}}{8} - \frac{\sqrt{3}}{8} e_1
\right )\\ & = - \frac{1}{8} e_4 - \frac{1}{8} e_2 +\frac{\sqrt{3}}{8} + \frac{\sqrt{3}}{8} e_1
\end{split}
\end{equation}
Next, we construct the anti-strange $V_{as}$ and anti-bottom $V_{ab}$, by left-multiplication of the anti-down quark $V_{ad}$ by $e^{2\pi e_3/3}$. 
\begin{equation}
\begin{split}
V_{as} &= \left(-\frac{1}{2} + \frac{\sqrt{3}}{2} e_3\right) \times V_{ad} =  \left(-\frac{1}{2} + \frac{\sqrt{3}}{2} e_3\right) \times \left( \frac{1}{4} e_5 + \frac{1}{4} e_3\right)\\
& = -\frac{1}{8} e_5 -\frac{1}{8} e_3 + \frac{\sqrt{3}}{8} e_2  - \frac{\sqrt{3}}{8}
\end{split}
\end{equation}
\begin{equation}
\begin{split}
V_{ab} & = \left(-\frac{1}{2} + \frac{\sqrt{3}}{2} e_3\right) \left(-\frac{1}{8} e_5 -\frac{1}{8} e_3 + \frac{\sqrt{3}}{8} e_2  - \frac{\sqrt{3}}{8}\right) \\
& = - \frac{1}{8}e_5 - \frac{\sqrt{3}}{8}e_2 - \frac{1}{8} e_3 +\frac{\sqrt{3}}{8}
\end{split}
\end{equation}
Next, we construct the octonions for the anti-muon $V_{a\mu}$ and anti-tau-lepton $V_{a\tau}$  by left multiplying the positron $V_{e^+}$ by $e^{2\pi e_1/3}$
\begin{equation}
\begin{split}
V_{a\mu} & =  \left(-\frac{1}{2} + \frac{\sqrt{3}}{2}e_1\right) \times \left( - \frac{1}{4}e_1 -\frac{1}{4} e_7\right) \\
& = \frac{1}{8} e_1 +\frac{1}{8} e_7 + \frac{\sqrt{3}}{8} + \frac{\sqrt{3}}{8} e_3
\end{split}
\end{equation} 
\begin{equation} 
\begin{split}
V_{a\tau}  &=   \left(-\frac{1}{2} + \frac{\sqrt{3}}{2}e_1\right)  \times \left( \frac{1}{8} e_1 +\frac{1}{8} e_7 + \frac{\sqrt{3}}{8} + \frac{\sqrt{3}}{8} e_3\right) \\
& = \frac{1}{8} e_7 - \frac{\sqrt{3}}{8} + \frac{1}{8} e_1 - \frac{\sqrt{3}}{8}e_3
\end{split}
\end{equation}
Lastly, we construct the octonions $V_{\nu\mu}$ for the muon neutrino and $V_{\nu\tau}$ for the tau neutrino, by left multiplying on the electron neutrino $V_{\nu}$ with $e^{2\pi e_6 /3}$ 
\begin{equation}
\left ( -\frac{1}{2} + \frac{\sqrt{3}}{2} {e_6} \right) \times \frac{1}{2} e_6 = -\frac{1}{4} e_6 - \frac{\sqrt{3}}{4}
\end{equation} 
\begin{equation} 
V_{\nu\tau} = \left ( -\frac{1}{2} + \frac{\sqrt{3}}{2} {e_6} \right) \times \left( -\frac{1}{4} e_6 - \frac{\sqrt{3}}{4}\right) =  -\frac{1}{4} e_6 + \frac{\sqrt{3}}{4}
\end{equation} 
We now have all the information needed to write down the four Jordan matrices whose eigenvalues we will calculate. 
Diagonal entries are electric charge, and off-diagonal entries are octonions representing the particles. Using the above results we write down these four matrices explicitly. 
The neutrinos of three generations
\begin{equation}
X_\nu =
\begin{bmatrix}
0 &  V_\nu & V_{\nu\mu}^*\\
V_\nu^* & 0 & V_{\nu\tau}\\
V_{\nu\mu} & V_{\nu\tau}^* & 0
\end{bmatrix}
\end{equation}
The anti-down set of quarks of three generations [anti-down, anti-strange, anti-bottom]: \begin{equation}
X_{ad} =
\begin{bmatrix}
\frac{1}{3} &  V_{ad} & V_{as}^*\\
V_{ad}^* & \frac{1}{3}& V_{ab}\\
V_{as} & V_{ab}^* & \frac{1}{3}
\end{bmatrix}
\end{equation}
The up set of quarks for three generations [up, charm, top]
\begin{equation}
X_u =
\begin{bmatrix}
\frac{2}{3} &  V_u & V_c^*\\
V_u^* & \frac{2}{3} & V_t\\
V_c & V_t^* & \frac{2}{3}
\end{bmatrix}
\end{equation}
The positively charged leptons of three generations [positron, anti-muon, anti-tau-lepton]
\begin{equation}
X_{e+}=
\begin{bmatrix}
1 &  V_{e+} & V_{a\mu}^*\\
V_{e+}^* & 1 & V_{a\tau}\\
V_{a\mu} & V_{a\tau}^{*} & 1
\end{bmatrix}
\end{equation}
Next, the eigenvalue equation corresponding to each of these Jordan matrices can be written down, after using the expressions given above for calculating the determinant and the function $S(X)$. Tedious but straightforward  calculations with the octonion algebra give the following four cubic equations:

Neutrinos: We get $Tr(X)=0, S(X) = -3/4, Det(X)=0$, and hence the cubic equation and roots
\begin{equation}
\lambda^3 - \frac{3}{4}\lambda = 0  \qquad ROOTS: \left( -\sqrt{2}\;\sqrt{\frac{3}{8}},\  0,  \  \sqrt{2}\; \sqrt{\frac{3}{8}}\right)
\label{nautreenos}
\end{equation}
Anti-down-quark + its higher generations [anti-down, anti-strange, anti-bottom]: We get $Tr(X)=1, S(X) = -1/24, Det (X) = -19/216 $, and the following cubic equation and roots
\begin{equation}
\begin{split}
\lambda^3 - \lambda^2 - \frac{1}{24}\lambda  + \frac{19}{216}  = 0\\
ROOTS: \  \frac{1}{3} - \sqrt{\frac{3}{8}},\  \frac{1}{3},\ \frac{1}{3} + \sqrt{\frac{3}{8}}
\end{split}
\label{jevad}
\end{equation}
Up quark + its higher generations [up, charm, top]: We get $Tr(X) = 2, S(X) = 23/24, Det(X) = 5/108$ and the following cubic equation and roots:
\begin{equation}
\begin{split}
\lambda^3 - 2 \lambda^2 + \frac{23}{24} \lambda - \frac{5}{108}  = 0\\
ROOTS :  \frac{2}{3} - \sqrt{\frac{3}{8}},\  \frac{2}{3}, \ \frac{2}{3} + \sqrt{\frac{3}{8}}
\end{split}
\end{equation}
Positron + its higher generations [positron, anti-muon, anti-tau-lepton]: We get $Tr(X) = 3, S(X) = 3-3/8, Det(X) = 1-3/8$ and the following cubic equation and roots:
\begin{equation}
\begin{split}
\lambda^3 - 3 \lambda^2 + \left(3 - \frac{3}{8}\right)  \lambda - \left( 1-\frac{3}{8}\right) = 0\\
ROOTS:  \ 1- \sqrt{\frac{3}{8}}, \  1, \ 1 + \sqrt{\frac{3}{8}}
\end{split}
\label{eleccu}
\end{equation}
As  expected from the known elementary properties of cubic equations, the sum of the roots is $Tr(X)$, their product is $Det(X)$, and the sum of their pairwise products is $S(X)$. Interestingly, this also shows that the sum of the roots is equal to the total electric charge of the three fermions under consideration in each of the respective cases. Whereas $S(X)$ and $Det(X)$ are respectively related to an invariant inner product and an invariant trilinear form constructed from the Jordan matrix, their physical interpretation in terms of fermion properties remains to be understood.

The roots exhibit a remarkable pattern. In each of the four cases, one of the three roots is equal to the corresponding electric charge, and the other two roots are placed symmetrically on both sides of the middle root, which is the one equal to the electric charge. All three roots are positive in the up quark set and in the positron set, whereas the neutrino set and anti-down quark set have one negative root each, and the neutrino also has a zero root. It is easily verified that the calculation of eigenvalues for the anti-particles yields the same set of eigenvalues, upto a sign. In other words, the Jordan eigenvalue for the anti-particle is opposite in sign to that for the particle. The roots are summarised in Fig. 3 below, and we see that they are composed of the electric charge, and the octonionic magnitude associated with the respective particle. [The octonionic magnitude $L_P^2 / L^2$ is the sum $\sum x_i x^i$ over the three identically charged fermions of three generations, which appears in Equation (5) above.]
 \begin{figure}[!htb]
        \center{\includegraphics[width=\textwidth]
        {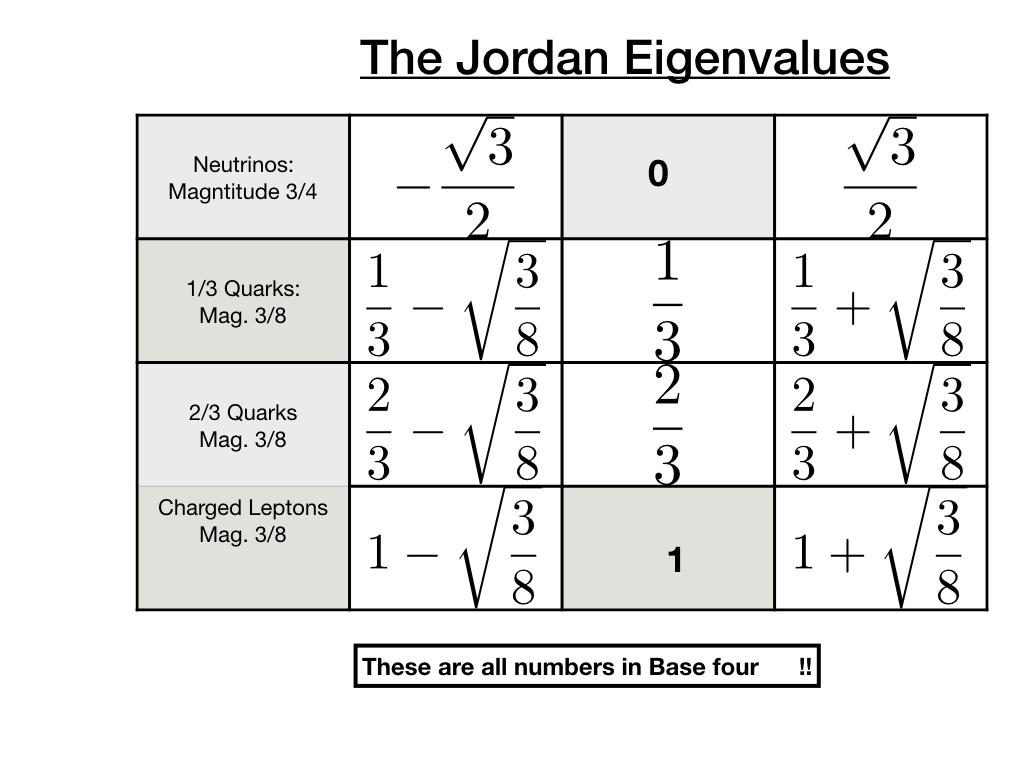}}
        \caption{\label{fig:my-label} The eigenvalues of the exceptional Jordan algebra for the various fermions. The eigenvalues are made from electric charge and the octonionic magnitude, and represent charge-mass of the corresponding fermion, in the pre-theory. The corresponding eigenmatrices \cite{Dray1} represent charge-mass eigenstates. The $SU(3)_c$ and $U(1)$ constructed from the $Cl(6)$ and the octonion algebra for one generation defines electric charge. However to define charge-mass and mass one must deal with $F_4$ and all three generations, not just one.}
      \end{figure}
One expects these roots to relate to masses of quarks and leptons for various reasons, and principally because the automorphism group of the complexified octonions contains the 4D Lorentz group as well, and the latter we know relates to gravity. Since mass is the source of gravity, we expect the Lorentz group to be involved in an essential way in any theory which predicts masses of elementary particles. And the group $F_4$, besides being related to $G_2$, and a possible candidate for the unification of the four interactions, is also the automorphism group of the EJA. We have motivated how the four projections of the EJA relate naturally to the four generation sets of the fermions. Thus there is a strong possibility that the eigenvalues of the characteristic equation of the EJA yield information about fermion mass ratios, especially it being a cubic equation with real roots. We make the following preliminary observations about the known mass ratios, and then provide a concrete analysis in Section IV.

The Jordan eigenvalues allow us to express the electric charge eigenstates of a fermion's three generations,  as superpositions of mass eigenstates. That is why these eigenvalues determine mass ratios.

For the set (positron, anti-muon, anti-tau-lepton), the three respective masses are known to satisfy the following empirical relation, known as the Koide formula:
\begin{equation}
\frac{ m_e + m_\mu + m_\tau}{( \sqrt{m_e} + \sqrt{m_\mu} + \sqrt{m_\tau})^2} = 0.666661(7) \approx \frac{2}{3}
\end{equation}
For the three roots of the corresponding cubic equation (\ref{eleccu}) we get that
\begin{equation}
2\frac{\lambda_1 ^2  +  \lambda_2 ^2 + \lambda_3 ^2}{ (\lambda_1 + \lambda_2 + \lambda_3)^2 } = 2\frac{[Tr(X)]^2-2S(X)}{[Tr(X)]^2} =\frac{2}{3} \left(1+\frac{1}{4}\right) \approx  0.8333
\end{equation}
The factor $1/4$ comes from the sum of the absolute values of the three octonions which go into the related Jordan matrix. This observation suggests that the eigenvalues bear some relation with the square roots of the masses of the three charged leptons, though simply comparing square roots of their mass-ratios does not seem to yield any obvious relation with the eigenvalues. Further investigation is presented in Section IV. Here, we observe the  following logarithmic ratios
for masses of  the charged leptons [taken as 0.5 MeV, 105 MeV, 1777 Mev] and for  the roots
\begin{equation}
\ln \left(  \frac{105}{0.5} \right)^{1/2} \sim 2.67; \qquad \frac{1}{1- \sqrt{\frac{3}{8}} }  \sim 2.58
\end{equation}
\begin{equation}
\ln \left(  \frac{1777}{0.5} \right)^{1/2} \sim 4.09; \qquad \frac{1+ \sqrt{\frac{3}{8}} } {1- \sqrt\frac{3}{8}} \sim 4.16
\end{equation}
\begin{equation}
\ln \left(  \frac{1777}{105} \right)^{1/2} \sim 1.41; \qquad \frac{1+ \sqrt{\frac{3}{8}} } {1} \sim 1.61
\end{equation}
For the up quark set though, we see a correlation in terms of square roots of masses.

In the case of the up quark set, the following approximate match is observed between the ratios of the eigenvalues, and the  mass square root ratios of the masses of up, charm and top quark. For the sake of this estimate we take these three quark masses to be [2.3, 1275, 173210] in Mev \cite{pdg}. The following ratios are observed:
\begin{equation}
\sqrt {  \frac{1275}{2.3}} \sim 23.55 ; \qquad \frac{ \frac{2}{3} + \sqrt{\frac{3}{8} }} {  \frac{2}{3} - \sqrt{\frac{3}{8} } }   \approx 23.56
\label{charmr}
\end{equation}
\begin{equation}
\sqrt{ \frac{173210}{1275}} \sim  11.66 ; \qquad  \frac{ \frac{2}{3}  } {  \frac{2}{3} - \sqrt{\frac{3}{8} } }  \approx 12.28
\end{equation}
\begin{equation}
\sqrt { \frac{173210}{2.3}} \sim  274.42 ; \qquad \left( \frac{ \frac{2}{3} + \sqrt{\frac{3}{8} }} {  \frac{2}{3} - \sqrt{\frac{3}{8} } }  \right) \times \left( \frac{ \frac{2}{3}  } {  \frac{2}{3} - \sqrt{\frac{3}{8} } }\right)  \approx 289.23
\label{topr}
\end{equation}
Within the error bars on the masses of the up set of quarks, the two sets of ratios are seen to agree with each other upto second decimal place.

Considering that one of the roots is negative in the anti-down-quark set, we cannot directly relate the eigenvalues to mass ratios.  The same is true for the neutrino set, where one root is negative and one root is zero. In section IV we propose that the correct quantity to examine is the square-root of mass (in dimensionless units), which can take both positive sign and negative sign: $\pm\sqrt{m}$. The Jordan eigenvalues relate to the square-root of either sign, with the eigenvalue for anti-particle being opposite in sign to that for the particle. The case of the neutrino is  especially instructive, and shows how non-zero mass could arise fundamentally, even when the electric charge is zero. In this case, the non-zero contribution comes from the inner product related quantity $S(X)$, and therein from the absolute magnitude of the octonions in the Jordan matrix, which necessarily has to be non-zero. We thus see that masses are derivative concepts, obtained from the three more fundamental entities, namely the electric charge, and the geometric invariants $S(X)$ and $Det(X)$,  with the last two necessarily being defined commonly for the three generations. And since mass is the source of gravity, this picture is consistent with gravity and space-time geometry being emergent from the underlying geometry of the octonionic space which algebraically determines the properties of the elementary particles. We note that there are no free parameters in the above analysis, no dimensional quantities, and no assumption has been put by hand. Except that we identify the octonions with elementary fermions. The numbers which come out from the above analysis are number-theoretic properties of the octonion algebra.

These observations suggest a possible fundamental relation between eigenvalues of the EJA and particle masses. In the next section, we provide further evidence for  such a connection, based on our proposal for unification based on division algebras and a matrix-valued Lagrangian dynamics.

\section{An octonionic  Lagrangian for the standard model} 
Our proposal is a matrix-valued Lagrangian dynamics, based on Adler's theory of trace dynamics. Adler  derives quantum field theory from a pre-quantum theory, where the latter is obtained by raising the classical dynamical degrees of freedom to the status of matrices/operators. However, the quantum Heisenberg algebra is not imposed a priori (instead, it is emergent). The Lagrangian thus becomes a matrix polynomial, and its matrix trace defines a scalar Lagrangian, whose integral over space-time volume yields the action, as usual. Variation of the action with respect to the matrix-valued configuration variables yields the matrix-valued Lagrange equations of motion from which Hamiltonian dynamics can be constructed in the usual way. Assuming that this dynamics holds at Planck time scale resolution, one asks as to what the coarse-grained emergent dynamics is, if one is observing the system at time scales much larger than Planck time, as in today's universe. Using the techniques of statistical thermodynamics, it is shown that the emergent dynamics is quantum field theory, with Planck's constant arising from the equipartition of a novel Noether charge unique to trace dynamics. In general the Hamiltonian of the trace dynamics theory is allowed to be have an anti-self-adjoint component, and quantum theory emerges only when this component is negligible. Critical entanglement amongst the matrix-valued configuration variables results in the anti-self-adjoint part of the Hamiltonian becoming significant, leading to a quantum-to-classical transition mediated by the Ghirardi-Rimini-Weber process of spontaneous localisation.

In trace dynamics, space-time is flat 4D Minkowski spacetime, and gravitation is not incorporated. Also, a fundamental trace Lagrangian for fermions and standard model gauge fields is not specified. In our generalisation of trace dynamics, we have proposed as to how gravity is to be incorporated, and what is the fundamental Lagrangian for gravity, and standard model fermions and gauge fields. The eigenvalues of the Dirac operator on a Riemannian geometry serve as dynamical variables of general relativity, and by raising these eigenvalues to the status of matrices/operators one arrives at the trace dynamical generalisation of classical gravitation - a pre-spacetime, pre-quantum theory. Furthermore, because gravitation is no longer classical here, the point structure of classical spacetime is lost. We replace it by the non-commuting coordinate geometry of the octonions, a spinor spacetime equivalent to an octonion-valued twistor geometry, which is equivalent to 10D Minkowski spacetime. Clifford algebras are used to define spinor states for fermions in this octonionic space, whose symmetries reveal the standard model, as well as its generalisation to a left-right symmetric extension of the standard model which includes pre-gravitation. The  exceptional Lie groups $G_2, F_4, E_6$ and $E_8$ all play a central role in the theory. Our theory is also the sought for reformulation of quantum theory which does not refer to classical time.

As in trace dynamics, a Lagrangian (described in the next sub-section) and an action principle is constructed. The Lagrangian describes a 2-brane on octonionic space [more strictly on split bioctonionic space] and we refer to this fundamental entity as an atom of space-time-matter [STM], or an `aikyon'. Our theory bears remarkable similarities to string theory (with pre-gravitation bearing a relation to loop quantum gravity) but also differs from string theory in crucial ways which help overcome the limitations of string theory. The universe is made of enormously many such STM atoms, an atom being nothing but an elementary particle along with all the fields it produces; thus the fermionic as well as bosonic aspect are unified into a common unit which is characterised by only one (length) parameter. Different particles are different vibrations of the 2-brane. As in trace dynamics, equations of motion are derived and a Hamiltonian dynamics constructed. Assuming the theory to hold at Planck time scale, a coarse-grained theory is derived [for $\tau\gg \tau_{Pl}$]. This emergent theory is relativistic weak quantum gravity; it is also a low-energy theory of unification, and the octonionic coordinate geometry is preserved, because coarse-graining is over the so-called Connes time which is extrinsic to octonionic space. Critical entanglement leads to the emergence of classical macroscopic density perturbations / objects, and the concurrent emergence of 4D classical  spacetime. However, quantum systems which are not critically entangled continue to live in the higher dimensional octonionic space which includes 4D spacetime as a subspace. The extra dimensions are complex-number valued and never compactified. These quantum systems when described by the generalised trace dynamics reveal the standard model and determine its free parameters. On the other hand, they can also be approximately described by quantum field theory on classical 4D spacetime with its point structure, but then the underlying octonionic space is lost, and we do not know why the standard model is what it is.

Trace dynamics was developed by Adler and collaborators in their papers \cite{Adler:94, AdlerMillard:1996} and is described in his book {\it `Quantum theory as an emergent phenomenon'} \cite{Adler:04}. Trace dynamics is also reviewed in summary accounts in \cite{RMP:2012, maithresh2019b, Singhreview}. Generalised trace dynamics is described in detail in \cite{Singhreview} where the construction of the Lagrangian to be described in the next subsection is detailed. In its most basic form (see Eqn. \ref{acnbasis} below), the action is for an STM atom prior to the Left-Right symmetry breaking, and has $E_8 \times E_8$ symmetry on complex split bioctonionic space. L-R symmetry breaking introduces the Yang-Mills coupling constant $\alpha$ as in Eqn. (\ref{fourthree}) below, and separates the left-handed electric charge particle eigenstates from right-handed square-root mass particle eigenstates. The relative amplitude coefficients of the LH states and the RH states cannot be arbitrary,  but are fixed by the octonion algebra. This is the precise reason why coupling constants are determined by the octonion geometry, even at low energies, via the fundamental action in its form in Eqn. (\ref{lagba}).

\subsection {A Lagrangian on an 8D octonionic space-time}
The action and Lagrangian for the three generations of standard model fermions, fourteen gauge bosons, and the  potential Higgs boson, are given by \cite{Singh2020DA}
\begin{equation}
\frac{S}{C_0} = \int d\tau\; {\cal L} \qquad  ; \qquad\mathcal{L} =  \frac{1}{2} Tr \biggl[\biggr. \dfrac{L_{p}^{2}}{L^{2}}  \dot{\widetilde{Q}}_{1}^{\dagger}\;  \dot {\widetilde{Q}}_{2} \biggr]
\label{acnbasis}
\end{equation}
Here,
\begin{equation}
\dot{\widetilde{Q}}_{1}^{\dagger}    =   \dot{\widetilde{Q}}_{B}^{\dagger} + \dfrac{L_{p}^{2}}{L^{2}} \beta_{1} \dot{\widetilde{Q}}_{F}^\dagger  ; \ \qquad \dot {\widetilde{Q}}_{2} =  \dot{\widetilde{Q}}_{B} + \dfrac{L_{p}^{2}}{L^{2}} \beta_{2} \dot{\widetilde{Q}}_{F}
\end{equation}
and 
\begin{equation}
{\dot{\widetilde{Q}}_B} = \frac{1}{L} (i\alpha q_B + L \dot{q}_B); \qquad  {\dot{\widetilde{Q}}_F} = \frac{1}{L} (i\alpha q_F + L \dot{q}_F) 
\label{fourthree}
\end{equation}
By defining
\begin{equation}
q^\dagger_1 = q_B^\dagger + \frac{L_P^2}{L^2} \beta_1 q^\dagger_F \qquad; \qquad q_2 = q_B + \frac{L_P^2}{L^2} \beta_2 q_F
\end{equation}
we can express the Lagrangian as
\begin{equation}
\begin{aligned}
{\cal L} &=  \frac{L_P^2}{2L^2} \; Tr \left[ \left(\dot{q}_1^\dagger + \frac{i\alpha}{L} q_1^\dagger \right) \times \left(\dot{q}_2 + \frac{i\alpha}{L} q_2 \right)\right]\\
&= \frac{L_P^2}{2L^2} \; Tr \left[  \dot{q}_1^\dagger \dot{q}_2 - \frac{\alpha^2}{L^2} q_1^\dagger q_2  + \frac{i\alpha}{L} q_1^\dagger \dot{q}_2  + \frac{i\alpha}{L} \dot{q}_1^\dagger q_2 \right]
\end{aligned}
\label{lagba}
\end{equation}
We now expand each of these four terms inside of the trace Lagrangian, using the definitions of $q_1$ and $q_2$ given above:
\begin{equation}
\begin{aligned}
\dot{q}_1^\dagger \dot{q}_2 & =  \dot{q}_{B}^\dagger \dot{q}_{B}  + \frac{L_P^2}{L^2} \dot{q}_B^\dagger \beta_2 \dot{q}_F + \frac{L_P^2}{L^2} \beta_1 \dot{q}_F^\dagger \dot{q}_B + \frac{L_P^4}{L^4} \beta_1 \dot{q}_F^\dagger \beta_2 \dot{q}_F \\
q_1^\dagger q_2 & = q_B^\dagger {q}_B + \frac{L_P^2}{L^2} q_B^\dagger \beta_2 {q}_F + \frac{L_P^2}{L^2} \beta_1 q_F^\dagger {q}_B + \frac{L_P^4}{L^4}\beta_1 q_F^\dagger  \beta_2 {q}_F \\
q_1^\dagger \dot{q}_2 & =   q_B^\dagger \dot{q}_B + \frac{L_P^2}{L^2} q_B^\dagger \beta_2 \dot{q}_F + \frac{L_P^2}{L^2} \beta_1 q_F^\dagger \dot{q}_B + \frac{L_P^4}{L^4}\beta_1 q_F^\dagger  \beta_2 \dot{q}_F \\
\dot{q}_1^\dagger q_2 & = \dot{q}_{B}^\dagger {q}_{B}  + \frac{L_P^2}{L^2} \dot{q}_B^\dagger \beta_2 {q}_F + \frac{L_P^2}{L^2} \beta_1 \dot{q}_F^\dagger {q}_B + \frac{L_P^4}{L^4} \beta_1 \dot{q}_F^\dagger \beta_2 {q}_F\\
\end{aligned}
\label{subequ}
\end{equation}
In our recent work, we suggested this Lagrangian, having the symmetry group $E_8$, as a candidate for unification. For the standard model sector, there are fourteen gauge bosons (equal to the number of generators of $G_2$). These are the eight gluons, the three weak isospin vector bosons, the photon, and the two Lorentz bosons. These bosons, along with one Higgs, can be accounted for by the four bosonic terms which form the first column in the above four sub-equations. The remaining twelve terms were proposed to describe three fermion generations and  Higgs, with the three generations being motivated by the triality of $SO(8)$. However, one important question which has not been addressed there is: why does triality not give rise to three copies of the bosons?! In the framework of the present approach we tentatively explore the following answer. We know that the even-grade Grassmann numbers which form the entries of the bosonic matrices are made from even-number products of odd-grade (fermionic) Grassmann numbers, and the latter are in a sense more basic. Could it then be that bosonic degrees of freedom are made from fermionic degrees of freedom? If this were to be so, it could prevent the tripling of bosons, if we think of them as arising at the `intersections' of the octonionic directions which represent fermions.

\subsection{An octonionic diagrammatic representation for three fermion generations, and fourteen gauge bosons, and the Higgs}
The seven imaginary unit octonions are used to make the Fano plane, which has seven points and seven lines [adding to fourteen elements; points and lines have equal status]. If we include the real direction [we have assumed $\dot{q}_{B0}$ to be self-adjoint] also, we get an equivalent of a 3-D cube where the eight vertices now stand for the eight octonions, with one of them [the `origin'] standing for the real line. As explained by Baez: ``The Fano plane is the projective plane over the 2-element field $Z_2$. In other words, it consists of lines through the origin in the vector space $Z_2^3$. Since every such line contains a single nonzero element, we can also think of the Fano plane as consisting of the seven nonzero elements of $Z_2^3$. If we think of the origin in $Z_2^3$ as corresponding to $1$ in ${\mathbb O}$, we get the following picture of the octonions". This picture is Fig. 4 below, borrowed from Baez \cite{baez2001octonions}.
 \begin{figure}[!htb]
        \center{\includegraphics[width=\textwidth]
        {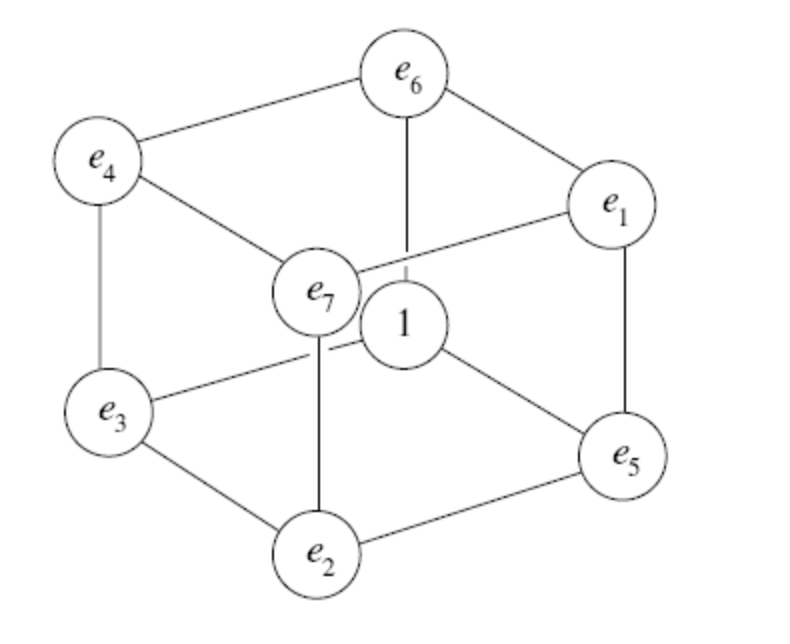}}
        \caption{\label{fig:my-label} The octonions [From Baez \cite{baez2001octonions}].}
      \end{figure}
Considering points, lines and faces together, this structure has 26 elements [8+12+6 = 26]. Motivated by this representation of the octonion, and the triality of $SO(8)$, we propose the following diagrammatic representation of the standard model fermions, gauge bosons, and Higgs as shown in Fig. 5. It motivates us to think of bosons as arising as `intersections' of the elements representing fermions. We have taken four copies of the Baez cube, with the central one at the intersection of the other three, and used them to represent the elementary particles. 
\begin{figure}[!htb]
        \center{\includegraphics[width=\textwidth]
        {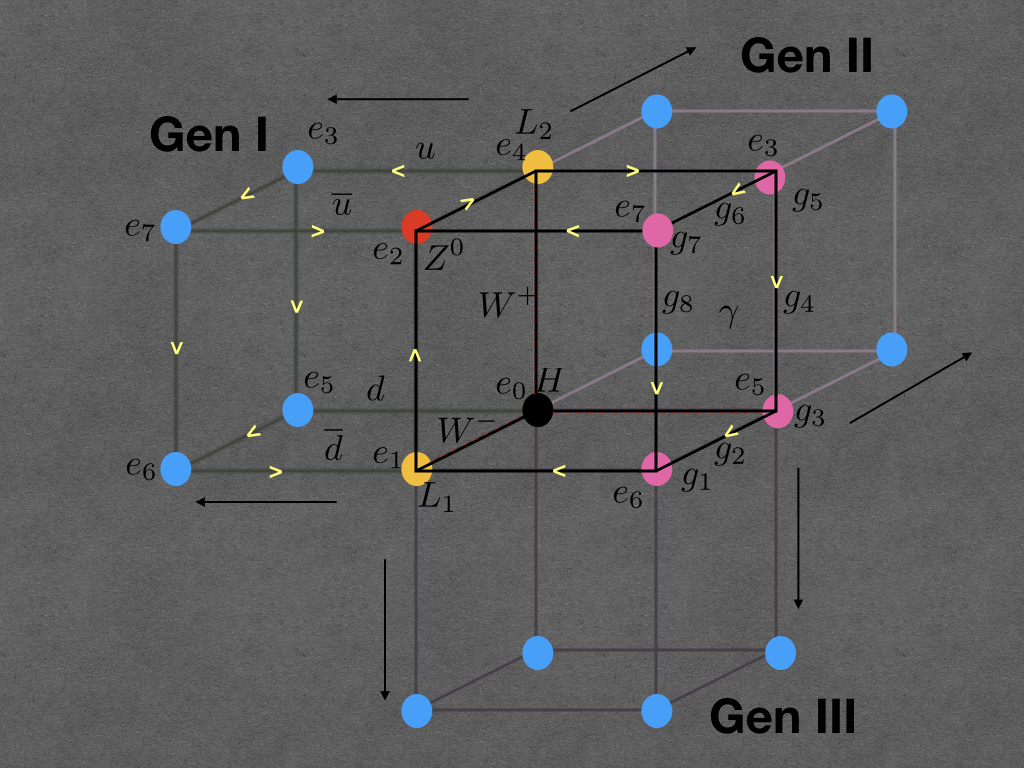}}
        \caption{\label{fig:my-label} The elementary particles of the standard model with three generations, represented through octonions in an $F_4$ diagram.. Please see text for a detailed explanation.}
      \end{figure}
We now attempt to describe Fig. 4 in some detail. There is a central black-colored cube (henceforth a cube is an octonion) in the front, which represents the fourteen gauge bosons and the four Higgs bosons; we will return to this cube shortly. Then there are three more (colored) cubes: one to the left, one at the back, and one at the bottom. These are marked as Gen I, Gen II and Gen III, and represent the three fermion generations. Let us focus first on the octonion on the left, which is Gen I, and where the eight vertices have been marked $(e_0, e_1, e_2, e_3, e_4, e_5, e_6, e_7)$ just as in the Baez cube. If $e_0$ were to be excluded, this cube becomes the Fano plane [Fig. 1 above] and the arrows marked in the Gen. I cube follow the same directions as in the Fano plane. In this Gen I cube, leaving out all those elements which are at the intersection with the central bosonic cube, and leaving out the face on the far left, we are left with sixteen elements: four points, eight lines, and four faces. The four points are shown in blue and are $(e_3, e_5, e_6, e_7)$. The eight lines are: $(e_4 e_3, e_7 e_2, e_3 e_7, e_7 e_6, e_5 e_6, e_6 e_4, e_5 e_0, e_6 e_1)$. The four planes are: $(e_4 e_3  e_7 e_2), (e_0 e_5 e_6 e_1), (e_7 e_2 e_1 e_6), (e_3 e_4 e_0 e_5)$. Between them, these sixteen elements represent the eight fermions and their anti-particles in one generation, one particle / anti-particle per octonionic element.

The up quark, the down quark, and their anti-particles of one particular color are (marked by) the four lines $(e_4 e_3, e_7 e_2, e_0 e_5, e_6 e_1)$. The points $(e_3, e_5, e_6, e_7)$ mark $u, d$ of a second color, and the lines $(e_3 e_7, e_7 e_6, e_3 e_5, e_5 e_6)$ mark the $u, d$ of the third color. The four planes mark the electron, the neutrino, and their anti-particles. Between them, these sixteen elements have an $SU(3)$ symmetry: they can be correlated to the (8+8)D particle basis constructed by Furey, from the $SU(3)$ in $G_2$. Next, the Gen II and Gen III along with Gen I has another $SU(3)$ symmetry, which is responsible for the three generations. These three fermionic cubes represent three intersecting copies of $G_2$ each cube having an $SU(3)$ symmetry. The three-way intersection is $SU(2)XSU(2)$, this being the black central cube, and the bosons lie on this cube. At the same time the fermionic cubes make contact with the bosonic cube, enabling the bosons to act on the fermions.

We now try to understand the central bosonic cube. First we count the number of its elements: it gets a total of 3x10=30 elements from the three side cubes, which when added to its own 26 elements gives a total of 56. But there are a lot of common elements, so that the actual number of independent elements is much smaller, and we enumerate them now. Three points are shared two-way and three points  shared three-way and the point $e_0$ is shared four-way; that reduces the count to 44. Nine lines are shared: three of them  three way, and six of them two way, reducing the count to 32. The shared three planes reduce the count to 29. We now account for the assignment of bosons to these 29 locations.

The eight gluons are on the front right, marked by the pink points, and lines labelled $g_1$ to $g_8$, and the photon is assigned to the plane $(e_3 e_7 e_6 e_5)$ on the front right enclosed by the gluons. The two Lorentz bosons are the yellow points $e_4$ and $e_1$ also marked $L_2$ and $L_1$. The three vector bosons are marked by the lines $e_0 e_1$, $e_0 e_4$ and the point $e_2$, also marked $Z^0$. The Higgs $H$ is at the four way real point  $e_0$. Three more Higgs are shown as follows: two planes per Higgs, e.g. the plane $e_0 e_ 4 e_2 e_1$ and the  mirror fermionic plane $e_3 e_5 e_6 e_7$ on the far left in Gen I. Analogously, another Higgs is given by the bosonic plane $e_0 e_1 e_6 e_5$ and its mirror fermionic plane at the front bottom in Gen III. The third Higgs is given by the bosonic plane $e_0 e_4 e_3 e_5$ and its mirror fermionic plane at the back in Gen II. This way 21 elements are used up. The remaining 8 un-used elements (six lines and two planes) are assigned to eight terms in the Lagrangian representing the action of the spacetime symmetry on the gluons: these are the terms $\dot{q}_B q_B^\dagger$ and $\dot{q}_B^\dagger q_B$ in (\ref{subequ}). 

The bosonic cube lies in the intersection of the three $G_2$ and hence does not triplicate during the $SU(3)$ rotation which generates the three fermion generations. The symmetry group of the theory is the 52 dimensional group $F_4$, with 8x3=24 generators coming from the three fermionic cubes, and the rest 28 from the bosonic sector [14 + 2x3 +  8 = 28]. This diagram does suggest that one could investigate bosonic degrees of freedom as made from pairs of fermion degrees of freedom. With this tentative motivation, we return to our Lagrangian, and seek to write it explicitly as for a single generation of bosons, and three generations of fermions. Upon examination of the sub-equations in Eqn. (\ref{subequ}) we find that the last column has terms bilinear in the fermions, and we would like to make it appear just as the second and third column do, so that we can explicitly have three fermion generations. With this intent, we propose the following assumed definitions of the bosonic degrees of freedom, by recasting the four terms in the last column of Eqn. (\ref{subequ}): 
\begin{equation}
\begin{split}
\frac{L_P^4}{L^4} \beta_1 \dot{q}_F^\dagger \beta_2 \dot{q}_F \equiv    \frac{L_P^2}{L^2} \dot{q}_B \beta_2 \dot{q}_F   + \frac{\alpha^2}{L^2}A  \\
\frac{L_P^4}{L^4}\beta_1 q_F^\dagger  \beta_2 {q}_F \equiv     \frac{L_P^2}{L^2} {q}_B \beta_2 {q}_F  + A \\
 \frac{L_P^4}{L^4}\beta_1 q_F^\dagger  \beta_2 \dot{q}_F \equiv \frac{L_P^2}{L^2} {q}_B^\dagger \beta_1 \dot{q}_F^\dagger     +B\\
 \frac{L_P^4}{L^4} \beta_1 \dot{q}_F^\dagger \beta_2 {q}_F    \equiv   \frac{L_P^2}{L^2}\dot{q}_B^\dagger \beta_1 q_F^\dagger  -B \\
\end{split}
\end{equation}
where $A$ and $B$ are  bosonic matrices which drop out on summing the various terms to get the full Lagrangian, With this redefinition, the sub-equations Eqn. (\ref{subequ}) can be now written in the following form after rewriting the last column:
\begin{equation}
\begin{aligned}
\dot{q}_1^\dagger \dot{q}_2 & =  \dot{q}_{B}^\dagger \dot{q}_{B}  + \frac{L_P^2}{L^2} \dot{q}_B^\dagger \beta_2 \dot{q}_F + \frac{L_P^2}{L^2} \beta_1 \dot{q}_F^\dagger \dot{q}_B +  \frac{L_P^2}{L^2} \dot{q}_B \beta_2 \dot{q}_F\\
q_1^\dagger q_2 & = q_B^\dagger {q}_B + \frac{L_P^2}{L^2} q_B^\dagger \beta_2 {q}_F + \frac{L_P^2}{L^2} \beta_1 q_F^\dagger {q}_B +\frac{L_P^2}{L^2} {q}_B \beta_2 {q}_F  \\
q_1^\dagger \dot{q}_2 & =   q_B^\dagger \dot{q}_B + \frac{L_P^2}{L^2} q_B^\dagger \beta_2 \dot{q}_F + \frac{L_P^2}{L^2} \beta_1 q_F^\dagger \dot{q}_B +  \frac{L_P^2}{L^2} {q}_B^\dagger \beta_1 \dot{q}_F^\dagger   \\
\dot{q}_1^\dagger q_2 & = \dot{q}_{B}^\dagger {q}_{B}  + \frac{L_P^2}{L^2} \dot{q}_B^\dagger \beta_2 {q}_F + \frac{L_P^2}{L^2} \beta_1 \dot{q}_F^\dagger {q}_B +   \frac{L_P^2}{L^2}\dot{q}_B^\dagger \beta_1 q_F^\dagger\\
\end{aligned}
\label{subeq}
\end{equation}
The terms now look harmonious and we can see a structure emerging - the first column are bosonic terms and these are not triples. The remaining terms are four sets of three each [to which their adjoints will eventually get added] which can clearly describe three generations of the four sets, which is what we had in the Jordan matrices in the previous section. Putting it all together, we can now rewrite the Lagrangian so that it explicitly looks like the one for gauge bosons and four sets of three generations of fermions, as in the Jordan matrix:
\begin{equation}
\begin{aligned}
{\cal L} &=  \frac{L_P^2}{2L^2} \; Tr \left[ \left(\dot{q}_1^\dagger + \frac{i\alpha}{L} q_1^\dagger \right) \times \left(\dot{q}_2 + \frac{i\alpha}{L} q_2 \right)\right]\\
&= \frac{L_P^2}{2L^2} \; Tr \left[  \dot{q}_1^\dagger \dot{q}_2 - \frac{\alpha^2}{L^2} q_1^\dagger q_2  + \frac{i\alpha}{L} q_1^\dagger \dot{q}_2  + \frac{i\alpha}{L} \dot{q}_1^\dagger q_2 \right]\\  \qquad \qquad \qquad \equiv \frac{L_P^2}{2L^2} \; & Tr \left[ {\cal L}_{bosons} + {\cal L}_{set1} + {\cal  L}_{set2} + {\cal L}_{set3} +{\cal L}_{set4} \right]
\end{aligned}
\label{lagfundam}
\end{equation}
where 
\begin{equation}
{\cal L}_{bosons} =  \dot{q}_{B}^\dagger \dot{q}_{B}  -\frac{\alpha^2}{L^2}  q_B^\dagger {q}_B +\frac{i\alpha}{L}  q_B^\dagger \dot{q}_B + \frac{i\alpha}{L}  \dot{q}_{B}^\dagger {q}_{B}
\end{equation}
\begin{equation}
{\cal L}_{set1} =  \frac{L_P^2}{L^2} \dot{q}_B^\dagger \beta_2 \dot{q}_F + \frac{L_P^2}{L^2} \beta_1 \dot{q}_F^\dagger \dot{q}_B +  \frac{L_P^2}{L^2} \dot{q}_B \beta_2 \dot{q}_F
\label{set1}
\end{equation}
\begin{equation}
{\cal L}_{set2} = -\frac{\alpha^2}{L^2} \left(\frac{L_P^2}{L^2} q_B^\dagger \beta_2 {q}_F + \frac{L_P^2}{L^2} \beta_1 q_F^\dagger {q}_B +\frac{L_P^2}{L^2} {q}_B \beta_2 {q}_F\right)
\label{chargedl}
\end{equation}
\begin{equation}
{\cal L}_{set3} = \frac{i\alpha}{L} \left(  \frac{L_P^2}{L^2} q_B^\dagger \beta_2 \dot{q}_F + \frac{L_P^2}{L^2} \beta_1 q_F^\dagger \dot{q}_B +  \frac{L_P^2}{L^2} {q}_B^\dagger \beta_1 \dot{q}_F^\dagger  \right) 
\label{qa1}
\end{equation}
\begin{equation}
{\cal L}_{set4} =  \frac{i\alpha}{L} \left(\frac{L_P^2}{L^2} \dot{q}_B^\dagger \beta_2 {q}_F + \frac{L_P^2}{L^2} \beta_1 \dot{q}_F^\dagger {q}_B +   \frac{L_P^2}{L^2}\dot{q}_B^\dagger \beta_1 q_F^\dagger\right)
\label{qa2}
\end{equation}
We see that each of these four fermionic sets could possibly be related to a Jordan matrix, after including the adjoint part. We also see that different coupling constants appear in different sets with identical coupling in third and fourth set and no coupling in the first set. The first set could possibly describe neutrinos, charged leptons and quarks (gravitational and weak interaction), the second set charged leptons and quarks, and the third and fourth set the quarks. To establish this explicitly, equations of motion remain to be worked out and then related to the eigenvalue problem. As noted earlier, $L$ relates to mass, and this approach could reveal how the eigenvalues of the EJA characteristic equation relate to mass. This investigation is currently in progress, and proceeds along the following lines. We take the self-adjoint part of the above Lagrangian, because that part is the one which leads to quantum field theory in the emergent approximation after coarse-graining the underlying theory. [The anti-self-adjoint part is negligible in the approximation in which quantum field theory emerges, and when it becomes significant, spontaneous localisation occurs, and classical space-time and the macroscopic universe emerges]. We vary the self-adjoint part of the Lagrangian with respect to the bosonic degree of freedom, and with respect to the three 8D-fermionic degrees of freedom, representing the three fermion generations. This yields four equations of motion, three of which are coupled matrix-valued Dirac equations for the three generations. These three coupled equations are solved by a state vector which is a three-vector made of three 8-spinors. The eigenvalue problem for  three coupled matrix equations is likely solved by the exceptional Jordan algebra, the algebra of 3x3 Hermitean matrices with octonionic entries, where the diagonal entries are identified with electric charge. That the diagonal entries are electric charge is justified  by the form of the Lagrangian above, especially as written in Eqn. (45), because we see $\alpha / L$ as the coefficient of the potential, and its square appearing in the electrodynamics term (52) in this latest form of the Lagrangian above. This coefficient in front of the terms in Eqn. (52) then gets identified with the fine structure constant, as below.

\bigskip

\subsection{The Jordan  eigenvalues and  the low energy limiting value of the fine structure constant }
If we examine the Lagrangian term for the charged leptons in Eqn. (\ref{chargedl}), the dimensionless coupiing constant $C$  in front of it is (upto a sign): 
\begin{equation}
C \equiv \alpha ^2 \frac{L_P^4}{L^4}
\end{equation}
[The operator terms of the form $q_B q_F$ etc. in (\ref{chargedl}) have been  correspondingly made dimensionless by dividing by $L_P^2$]. We assume that $\ln \alpha$ is linearly proportional to the electric charge, and that the proportionality constant is the Jordan eigenvalue corresponding to the anti-down quark. The electric charge $1/3$ of the anti-down quark seems to be the right choice for determining $\alpha$, it being the smallest non-zero value [and hence possibly the  fundamental value]  of the electric charge, and also because the constant $\alpha$ appears as the coupling in front of the supposed quark terms in the Lagrangian, as in Eqns. (\ref{qa1}) and (\ref{qa2}). We hence define $\alpha$ by
\begin{equation}
\ln \alpha \equiv \lambda_{ad}\;  q_{ad} = \left [ \frac{1}{3} - \sqrt\frac{3}{8} \right ] \times \frac{1}{3} \qquad \implies \quad \alpha^2 \approx = 0.83025195149
\label{alph}
\end{equation}
where $\lambda_{ad}$ is the Jordan eigenvalue corresponding to the anti-down quark, as given by Eqn. (\ref{jevad}) and $q_{ad}$ is the electric charge of the anti-down quark (=1/3). In order to arrive at this relation for $\alpha$, we asked in what way $\alpha$ could vary  with $q$, if it was allowed to vary? We then made the  assumption that $d\alpha/dq \propto \alpha$. In the resulting linear dependence of $\ln\alpha$ on $q$, we froze the value of $\alpha$ at that given by the smallest non-zero charge value $1/3$, taking the proportionality constant to be the corresponding Jordan eigenvalue. This dependence also justifies that had we fixed $\alpha$ from the zero charge of the neutrino, $\alpha$ would have been one, as it in fact is, in our Lagrangian.   We have justified this way of constructing $\alpha$ from the Lagrangian dynamics in \cite{vvs, Singhreview}, on the basis of a Left-Right symmetry breaking which results in the emergence of the standard model, and pre-gravitation, from the unbroken symmetry group $E_6$. Briefly put, prior to symmetry breaking the theory is scale invariant [only an overall length parameter in front of the trace Lagrangian], and the coupling constant $\alpha$ arises as a result of the symmetry breaking, as characterised by Eqn. (\ref{fourthree}), reproduced below:
\begin{equation}
{\dot{\widetilde{Q}}_B} \rightarrow \frac{1}{L} (i\alpha q_B + L \dot{q}_B); \qquad  {\dot{\widetilde{Q}}_F} \rightarrow \frac{1}{L} (i\alpha q_F + L \dot{q}_F) 
\label{fourfour}
\end{equation}
A `square-root-mass-charge' quantum number having an eigenvalue $1/3$ in the unified theory prior to symmetry breaking and associated with the algebra of split bioctonions $O\times O'$ also splits as follows, with the LH eigenvalue $1/3\times (1/3-\sqrt{3/8})$ becoming related to the electric charge of the down quark [and hence determining $\ln \alpha$] and the RH eigenvalue $1/3\times(1/3+\sqrt{3/8})$ becoming related to the square root of mass of electron:
\begin{equation}
\frac{1}{3} O\times O' \rightarrow  \frac{1}{2}\left\{1/3\times (1/3-\sqrt{3/8})\; O \oplus 1/3\times(1/3+\sqrt{3/8})\; O'\right\}
\end{equation}
This same split is also the reason why the mass ratios [as proposed below in the paper] arise the way they do. 

As for the value of $L_P / L$, we identify it with one-half of that part of the Jordan eigenvalue which modifies the contribution coming from the electric charge. 
[For an explanation of the origin of the factor of one-half, see the next paragraph]. 
Thus from the eigenvalues found above, we deduce that for neutrinos, quarks and charged leptons, the quantity $L_P^2 / L^2$ takes the respective values $(3/16, 3/32, 3/32)$. These values are equal to one-fourth of the respective octonionic magnitudes. Thus the coupling constant $C$ defined above can now be calculated, with $\alpha^2$ as given above, and $L_P^2/L^2 = 3/32$. Furthermore, since the electric charge $q$, the way it is conventionally  defined, has dimensions such that $q^2$ has dimensions (Energy $\times$ Length), we measure $q^2$ in Planck units $E_{Pl} \times L_P = \hbar c$. We hence  define the fine structure constant  by $C=\alpha^2 L_P^4 / L^4\equiv e^2 /\hbar c$, where $e$ is the electric charge of electron / muon / tau-lepton in conventional units. We hence get the value of the fine structure constant to be
\begin{equation}
C=\alpha^2 L_P^4 / L^4\equiv e^2 /\hbar c =  \exp \left[ \left [ \frac{1}{3} - \sqrt\frac{3}{8} \right ] \times \frac{2}{3} \right] \times \frac{9}{1024} \approx  0.00729713 = \frac{1}{137.04006}
\label{fsc}
\end{equation}
 The CODATA 2018 value of the fine structure constant is
 \begin{equation}
 0.0072973525693(11) = 1/137.035999084(21)
 \end{equation}
  Our calculated value differs from the measured value in the seventh decimal place. In the next section, we show how incorporating the Karolyhazy length correction gives an exact match with the CODATA 2018 value, if we assume a specific value for the electro-weak symmetry breaking energy scale.
  
  Why did we identify $L_P/L$ with one-half of the octonionic magnitude $\sqrt{3/8}$ rather than with the magnitude $\sqrt{3/8}$ itself? 
  The answer lies in the physical interpretation originally assigned to the length scale $L$. [Please see the discussion below Eqn. (69) of \cite{maithresh2019}]. The length $L$ for an object of mass $m$ is interpreted as the Schwarzschild radius $2Gm/c^2$ of an object of mass $m$, so that $L_P/L=L_P\;c^2/2Gm$, which is one-half the Compton wave-length (in Planck units)  and not the Compton wavelength itself. Assuming that the octonionic magnitude has to be identified with Compton wavelength (in units of Planck length), it hence has to be divided by one-half, before equating it to $L_P/L$. This justifies taking $L_P^2 / L^2 = 1/4 \times 3/8 = 3/32$. 
      
  Once a theoretical derivation of the asymptotic fine structure constant is known, one can write the electric charge $e$ as
\begin{equation}
e = (3/32) \exp [ 1/9 - 1/\sqrt{24} ] \ ( \hbar L_P / t_P )^{1/2}
\label{echarge}
\end{equation}
where $L_P$ and $t_P$ are Planck length and Planck time respectively  - obviously their ratio is the speed of light.
In our theory, there are only three fundamental dimensionful quantities: Planck length, Planck time, and a constant with dimensions of action, which in the emergent quantum theory is identified with Planck's constant $\hbar$.
We now see that electric charge is not independent of these three fundamental dimensionful constants. It follows from them.
Planck mass is also constructed from these three, and electron mass will be expressed in terms of Planck mass, if only we could understand why the electron is some $10^{22}$ times lighter than Planck mass. Such a small number cannot come from the octonion algebra. In all likelihood, the cosmological expansion up until the electroweak symmetry breaking is playing a role here.

Thus electric charge and mass can both be expressed in terms of Planck's constant, Planck length and Planck time. This encourages us to think of electromagnetism, as well the other internal symmetries, entirely in geometric terms. This geometry is dictated by the $F_4$ symmetry of the exceptional Jordan algebra.

\subsection {The Karolyhazy correction to the asymptotic value of  fine structure constant}  In accordance with the Karolyhazy uncertainty relation (Eqn. (9) of \cite{Singh:KL})  a measured length $l$ has a `quantum gravitational' correction $\Delta l$ given by
\begin{equation}
(\Delta l)^3 = L_P^2 \; l
\end{equation} 
For the purpose of the present discussion we shall assume an equality sign here, i.e. that the numerical constant of proportionality between the two sides of the equation is unity. And, for the sake of the present application to the fine structure constant, we rewrite this relation as
\begin{equation}
\delta \equiv \frac{L_P}{\Delta l} = \left(\frac{L_P}{l}\right)^{1/3}
\end{equation}
We set $l\equiv l_f$ where $l_f$ is the length scale ($\approx 10^{-16}$ cm) associated with electro-weak symmetry breaking, where classical space-time emerges from the prespacetime, prequantum theory. The assumption being that when the universe evolves from the Planck scale to the electro-weak scale [while remaining in the unbroken symmetry phase], the 
inverse of the octonionic length associated with the charged leptons (this being $\sqrt{3/32}$) is reset, because of the Karolyhazy correction, to
\begin{equation}
\sqrt\frac{3}{32} \qquad \longrightarrow \sqrt\frac{3}{32} + \delta_f \equiv \sqrt\frac{3}{32} + \left(\frac{L_P}{l_f}\right)^{1/3} 
\end{equation} 
We can also infer this corrected length as the four-dimensional space-time measure of the length, which differs from the eight dimensional octonionic value $\sqrt{3/32}$  by the amount $\delta_ f$. If we take $l_f$ to be $10^{-16}$ cm, the correction $\delta_f$ is of the order $2\times 10^{-6}$. The correction to the asymptotic value (\ref{fsc}) of the fine structure constant is then
\begin{equation}
C = \alpha^{2} L_P^4 /L^4 \equiv e^2/\hbar c = \alpha^2 \left [ \sqrt\frac{3}{32} + \left(\frac{L_P}{l_f}\right)^{1/3} \right]^4
\end{equation}
For $l_f = 10^{-16}$ cm =  198 GeV$^{-1}$, we get the corrected value of the fine structure constant to be $0.00729737649$, which overshoots the measured CODATA 2018 value at the eighth decimal place. The electroweak scale is generally assumed to lie in the range 100 - 1000 GeV. The value $l_f = 1.3699526 \times 10^{-16}$ cm =  144.530543605 GeV$^{-1}$ reproduces the CODATA 2018 value 0.0072973525693 of the asymptotic fine structure constant. The choice $l_f^{-1} = 246$ GeV gives the value 0.00729739452, whereas the choice
$l_f^{-1} = 159.5 \pm 1.5$ GeV gives the range (0.00729736049, 0.00729735908). 100 GeV gives the value 0.00729732757 which is smaller than the measured value.  1000 GeV gives 0.00729754842. Thus in the entire 100 - 1000 GeV range, the derived constant agrees with the measured value at least to the sixth decimal place, which is reassuring. The purpose of the present exercise is to show that the Karolyhazy correction leads to a correction to the asymptotic value of the fine structure constant which is in the desired range - a striking fact by itself. In principle, our theory should predict the precise value of the electroweak symmetry breaking scale. Since that analysis has not yet been carried out, we predict that the ColorElectro-WeakLorentz symmetry breaking scale is 144.something GeV, because only then the theoretically calculated value of the asymptotic fine structure constant matches the experimentally measured value.

The above discussion of the asymptotic low energy value of the fine structure constant should not be confused with the running of the constant with energy. Once we recover classical spacetime and quantum field theory from our theory, after the ColorElectro-WeakLorentz symmetry breaking, conventional RG arguments apply, and the running of couplings with energy is to be worked out as is done conventionally. Such an analysis of running couplings will however be valid only up until the broken symmnetry is restored - it is not applicable in the prespacetime prequantum phase. In this sense, our theory is different from GUTs. Once there is unification, Lorentz symmetry is unified with internal symmetries - the exact energy scale at which that happens remains to be worked out.

How then does the Planck scale prespacetime, prequantum theory know about the low energy asymptotic value of the fine structure constant? The answer to this question lies in the Lagrangian given in (\ref{lagfundam}) and in particular the Lagrangian term (\ref{chargedl}) for the charged leptons. In determining the asymptotic fine structure constant from here, we have neglected the  modification to the coupling that will come from the presence of $q_B$ and $q_F$. This is analogous to examining the asymptotic, flat spacetime limit of a spacetime geometry due to a source - gravity is evident close to the source, but hardly so, far from it. Similarly, there is a Minkowski-flat analog of the octonionic space, wherein the effect of $q_B$ and $q_F$ (which in effect `curve' the octonionic space) is ignorable, and the asymptotic fine structure constant can be computed. The significance of the non-commutative, non-associative octonion algebra and the Jordan eigenvalues lies in that they already determine the coupling constants, including their asymptotic values. This is a property of the algebra, even though the interpretation of a particular constant as the fine structure constant comes from the dynamics, i.e. the Lagrangian, as it should, on physical grounds.

The pre-quantum, pre-spacetime  theory is needed  both in UV and in IR, and the octonionic theory is such a theory. UV is obvious, but IR needs justification.
As explained in the Appendix, even at low energies, there can be a situation where for a given system, all sub-systems have action of the order $\hbar$. Then there is no background classical spacetime anymore, and the pre-quantum, pre-spacetime theory is required. e.g. when a massive object is in a quantum superposition of two position states and we want to know what spacetime geometry  it produces.
The pre-theory is in principle required also for a more exact description and understanding of the standard model, even at low energies. And the octonionic theory achieves just that, thereby being able to derive the low-energy SM parameters. This is BSM in IR, and has implications for how we plan BSM experiments: these have to be not only towards UV, but also in the IR.
The O-theory has only three fundamental constants, and these happen to be such that both the UV and IR limits can be easily investigated. These constants are Planck length, Planck time and Planck's constant $\hbar$. Note that, as compared to conventional approaches to quantum gravity, Planck's constant $\hbar$ has been traded for Planck mass/energy. And this is very important:
The pre-theory is in principle required whenever one or more of the following three conditions are satisfied: times scales $T$ of interest are order Planck time, Length scales $L$ of interest are order Planck length, actions S of interest are order $\hbar$.
If $T$ and $L$ are respectively   much larger than Planck time and Planck length, but S is of order $\hbar$, that requires the pre-theory in IR. 
If $T$ is order Planck time, then the energy scale $\hbar / T$ is  Planck  energy scale. However, $\hbar / T$ is in IR for $T \gg$ Planck time, and yet the pre-theory is required (for an exact in-principle description of SM) if all actions are order $\hbar$.
The BSM physics in IR is achieved by replacing 4D Minkowski spacetime by 8D octonionic non-commutative spacetime. This is the pre-theory analog of flat spacetime - and it has consequences - it predicts the low energy SM parameters, without switching on high-energy interactions in the UV.
Going to high energies is just like in GR. In GR we switch on the gravitational field around Minkowski spacetime and doing so takes us from IR to UV. Same way, in the O-theory we switch on SM interactions and would-be-gravity, *around* the `flat' octonionic spacetime [=10D Minkowski] and this takes us from IR to UV. But unlike in the GR case, we already learn a lot of BSM physics in the IR, because the spacetime is non-commutative. String Theory missed out on this important IR physics, because it continued to work with 10D Minkowski spacetime which is commutative, and from there went to UV. One should have looked at octonionic spacetime and Clifford algebras.

The importance and significance of the IR limit of the pre-theory is brought out in Fig. 6 below. The ground state of the theory is not Minkowski space-time [shown in the left of the figure] but the non-commutative generalisation of Minkowski spacetime [shown in the right of the figure]. 
\begin{figure}[!htb]
        \center{\includegraphics[width=\textwidth]
        {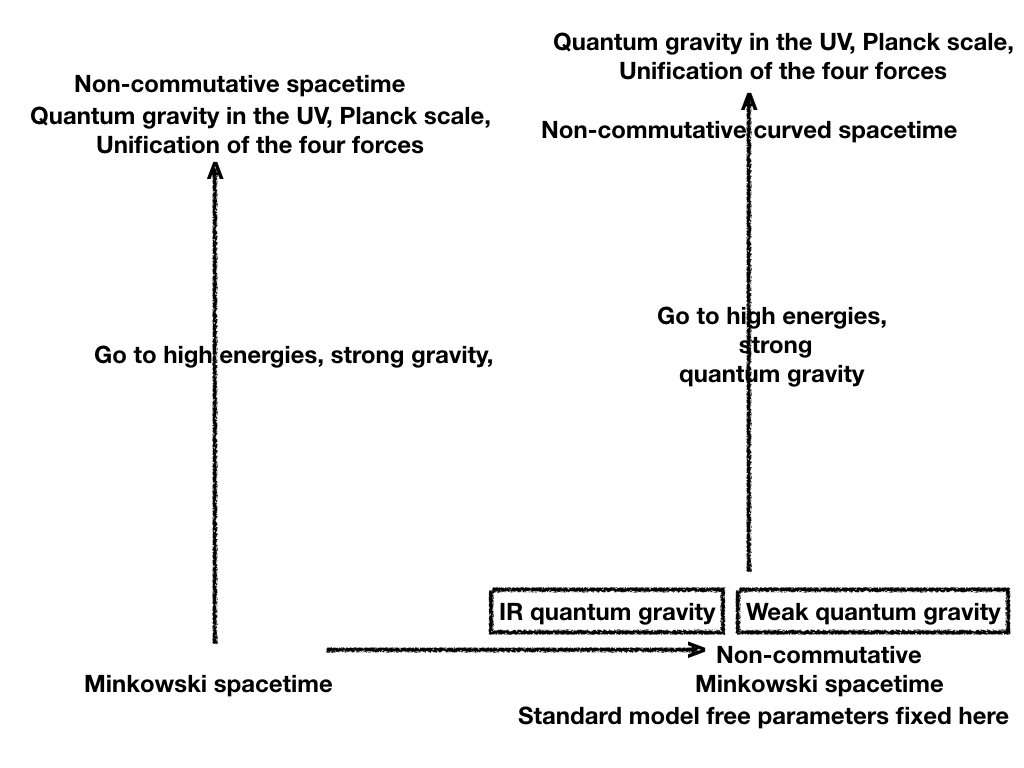}}
        \caption{\label{fig:my-label} The significance of the IR limit of the pre-quantum, pre-spacetime theory. Standard model free parameters are fixed in the IR, not in the UV region}
      \end{figure}

\subsection{Comparison with, and improvement on, string theory}

When it was proposed that elementary particles are not point objects, but extended like strings, one important conceptual issue was not clear. Why strings? What is the foundational principle / symmetry principle which compels us to consider extended objects?
In our work we started by demanding that there ought to exist a reformulation of quantum field theory which does not depend on classical time. This is the starting foundational question.
The symmetry principle then emerged as: physical laws must be invariant under general coordinate transformations of {\it non-commuting} coordinates.
The Lagrangian dynamics which implements these requirements requires (for consistency) that elementary particles be described by extended objects (strings). And it also requires that the theory be formulated in 10 spacetime dimensions, plus Connes time as an absolute time parameter.

This essentially is (an improved version of) string theory, with important accompanying changes:

(i) The vacuum is not 10D Minkowski vacuum. It is the octonionic 8D  algebraic vacuum. This makes it easy to relate the theory to the standard model.
(ii) The Hamiltonian at the Planck scale is not self-adjoint. This permits compactification without compactification: 4D classical spacetime is recovered without curling up the extra six dimensions.

We thus provide a quantum foundational motivation for string theory. Also, the underlying dynamics is deterministic and non-unitary; thus bringing the deterministic and reduction aspect of quantum theory in one unified new dynamics. It was once said that when string theory is properly understood, the quantum measurement problem will solve itself. In a manner of speaking, that is now seen to be true!
We have a pre-quantum, pre-spacetime dynamics, from which quantum field theory is emergent. The original aspects of string theory which still remain are: elementary particles are extended objects (strings) and they live in 10D Minkowski spacetime. Hence the octonionic theory could also be called improved string theory, in a significantly improved and falsifiable versioin, which has predictive power, and which can be tested in the laboratory.  

\bigskip

On a related note about this approach to unification, we recall that the symmetry group in our theory is $U(1) \times SU(3) \times SU(2) \times SU(2)$. This bears resemblance to the study of a left-right symmetric extension of the standard model by Boyle \cite{BoyleLR} in the context of the complexified exceptional Jordan algebra. This $L-R$ model has exceptional phenomenological promise, and it appears that  the unbroken phase [prior to the ColorElectro-WeakLorentz symmetry breaking] of the L-R model is well-described by our Lagrangian (\ref{lagfundam}) for three generations. This gives further justification for exploring the phenomenology of this Lagrangian.

\subsection{Quantum Worlds vs. Classical Worlds}

Our universe, as it is today, is dominated by classical bodies, which produce, and live in, a classical spacetime. This is the substrate shown in Fig. 7 below.
There is a sprinkling of systems for which we get to see the actual quantum behaviour. These are shown at the top left of the figure.
Then we realise that classical systems are a limiting case of quantum systems. This is shown by the curved arrowhead on the left of the figure.
Our current formulation of quantum systems embeds them in a classical spacetime, as depicted by the coloured arrow marked `Approximate'. Howsoever successful by the standard of current experiments, this formulation can only be approximate. Because the truth is, we do not really need the classical substrate, nor the classical spacetime, to describe quantum systems.

Quantum systems produce and live in a quantum spacetime, as shown in the top right. This is where the standard model truly lives, irrespective of what energy scale we study. However we currently describe the standard model only approximately, which is why our understanding of the standard model is only partial.
Note that this diagram makes no reference to the energy scale. It is true at all energies. UV as well as IR. The octonionic theory achieves a formulation of  the upper quantum level, relating quantum systems to a quantum spacetime, i.e. the octonionic space-time.
\begin{figure}[!htb]
        \center{\includegraphics[width=\textwidth]
        {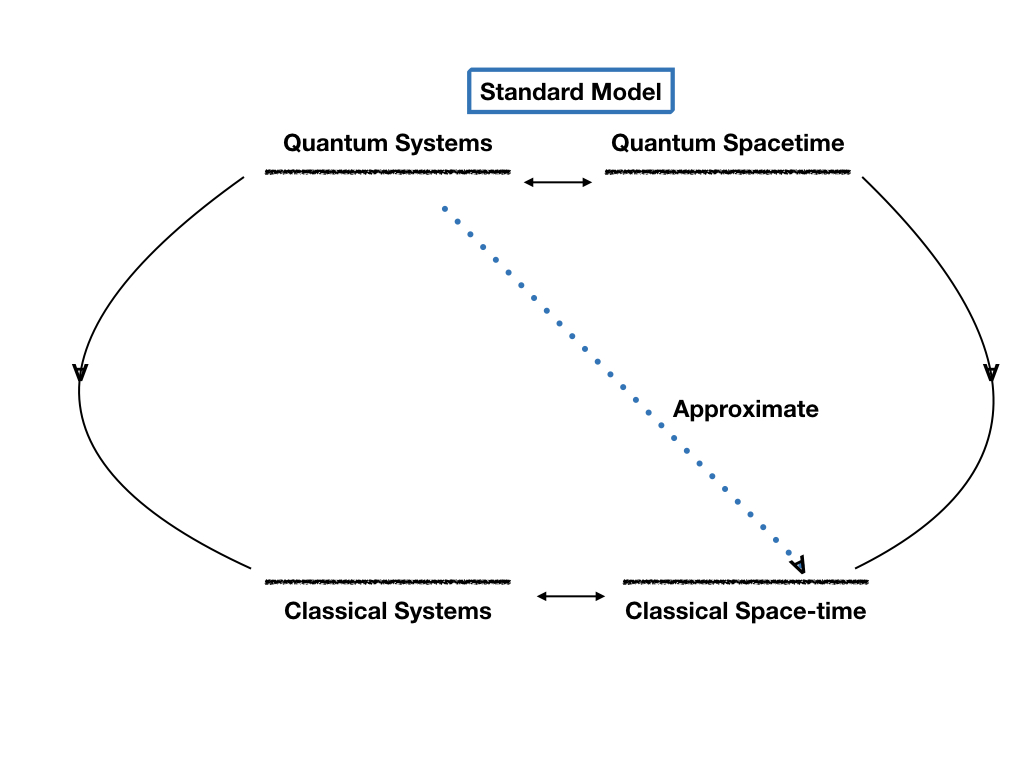}}
        \caption{\label{fig:my-label} Quantum world vs. classical world. The exact description of the standard model is at the upper level. Its current description is approximate.}
      \end{figure}

In the octonionic theory, the transition from the classical substrate to the upper quantum level is very elegant, and is reflected in the simplicity of the action principle of the theory. Let us recall from Eqn. (\ref{acnbasis}) and from Eqn. (53) of \cite{Singhreview} that the action principle of the theory is [an atom of space-time-matter (STM)]
\begin{equation}
 \frac{S}{\hbar} = \frac{1}{2}\int \frac{d\tau}{\tau_{Pl}}\; Tr \biggl[\biggr. \dfrac{L_{p}^{2}}{L^{2}}  \dot{\widetilde{Q}}_{1sed}^\dagger\;  \dot {\widetilde{Q}}_{2sed} \biggr]
\end{equation}
where the degrees of freedom are on sedenionic space. This action is nothing but a refined form of the action of a relativistic particle in curved  classical spacetime, i.e.
\begin{equation}
S = mc \int ds
\end{equation}
as we now explain. Consider the octonionic space and the matrix valued dynamical variable $q$ on it:
\begin{equation}
O = a_0 e_0 + a_1 e_1 + a_2 e_2 + a_3 e_3 + a_4 e_4 + a_5 e_5 + a_6 e_6 + a_7 e_7
\label{octad}
\end{equation}
where $e_0=1$, and the other $e_i$ are seven imaginary unit directions, each of which square to minus one, and obey the Fano plane multiplication rules. A dynamical variable, this being a matrix $q$ as in trace dynamics, will have eight component matrices; thus
\begin{equation}
q = q_0 e_0 + q_1 e_1 + q_2 e_2 + q_3 e_3 + q_4 e_4 + q_5 e_5 + q_6 e_6 + q_7 e_7
\label{matd}
\end{equation}
Eqn. (\ref{octad}) defines an octonion, whose eight direction vectors define the underlying physical space in which the `atom of space-time-matter' [the $Q$ matrices = elementary particles] lives.
The form of the matrix is shown in Eqn. (\ref{matd}). The elementary particles are defined by different directions of octonions. The $Q$-matrices as shown in the action define the `kinetic energy' of the STM atom. The trace is a matrix trace. Noting that $L$ is proportional to square root of mass $m$ (please refer next section)  the action can be written as
\begin{equation}
                                                                 S = \frac{1}{2}mc \int d\tau\; Tr  \biggl[\biggr. \dfrac{L_{p}^{2}}{L^{2}}  \dot{\widetilde{Q}}_{1sed}^\dagger\;  \dot {\widetilde{Q}}_{2sed} \biggr] 
 \end{equation}                                                                
Our fundamental action is a relativistic matrix-particle in higher dimensions.
The universe is made of enormously many such STM atoms which interact through `collisions' and entanglement. From their interactions emerges the low energy universe we see.
There perhaps cannot be a simpler description of unification than this action principle. Note that $Q_1$ and $Q_2$ are two different matrices which together define one `particle' hence giving it the character of an extended object such as the string of string theory.

Once again, we see the great importance of Connes time $\tau$. The universe is a higher dimensional spacetime manifold filled with matter, all evolving in an absolute Connes time.

\bigskip

\section{Jordan eigenvalues and mass-ratios}
We have not addressed the question as to how these discrete order one eigenvalues might relate to actual low values of fermion masses, which are much lower than Planck mass. We speculatively suggest the following scenario, which needs to be explored further. The universe is eight-dimensional, not four. The other four internal dimensions are not compactified; rather the universe is very `thin' in those dimensions but they are expanding as well. There are reasons having to do with the so-called Karolyhazy uncertainty relation \cite{Singh:KL}, because of which the universe expands in the internal dimensions at one-third the rate, on the logarithmic scale, compared to our 3D space. That is, if the 4D scale factor is $a(\tau)$, the internal scale factor is $a^{1/3}_{int}{(\tau})$, in Planck length units. Taking the size of the observed universe to be about $10^{61}$ Planck units, the internal dimensions have a width approximately $10^{20}$ Planck units, which is about $10^{-13}$ cm, thus being in the quantum domain. Classical systems have an internal dimension width much smaller than Planck length, and hence they effectively stay in [and appear to live in] four dimensional space-time. Quantum systems probe all eight dimensions, and hence live in an octonionic universe.

The universe began in a unified phase, via an inflationary 8D expansion possibly resulting as the aftermath of a huge spontaneous localisation event in a `sea of atoms of space-time-matter' \cite{maithresh2019}. The mass values are set, presumably in Planck scale, at order one values dictated by the eigenvalues reported in the present paper. Cosmic inflation scales down these mass values at the rate $a^{1/3}(\tau)$, where $a(\tau)$ is the 4D expansion rate. Inflation ends after about sixty e-folds, because seeding of classical structures breaks the color-elctro-weak-Lorentz symmetry, and classical spacetime emerges as a broken Lorentz symmetry. The electro-weak symmetry breaking is actually a electro-weakLorentz symmetry breaking, which is responsible for the emergence of gravity, weak interaction being its short distance limit. There is no reheating after inflation; rather inflation resets the Planck scale in the vicinity of the electro-weak scale, and the observed low fermion mass values result. The electro-weak symmetry breaking is mediated by the Lorentz symmetry, in a manner consistent with the conventional Higgs mechanism. It is not clear why inflation should end specifically at the electro-weak scale: this is likely dictated by when spontaneous localisation becomes significant enough for classical spacetime to emerge. It is a competition between the strength of the electro-colour interaction which attempts to bind the fermions, and the inflationary expansion which opposes this binding. Eventually, the expanding universe cools enough for spontaneous localisation to win, so that the Lorentz symmetry is broken. It remains to prove from first principles that this happens at around the electro-weak scale and also to investigate the possibly important role that Planck mass primordial black holes might play in the emergence of classical spacetime. I would like to thank Roberto Onofrio for correspondence which has influenced these ideas. See also \cite{Onofriodiscuss}.

\subsection{Evidence of correlation between the Jordan eigenvalues and the mass ratios of quarks and charged leptons} In the first generation, we note the positron mass to be $0.511$ Mev, the up quark mass to be $2.3 \pm 0.7 \pm 0.5$ MeV, and the down quark mass to be $4.8 \pm 0.5 \pm 0.3$ MeV. The  uncertainties in the two quark masses permit us to make the following proposal: the square-roots of the masses of the positron, up quark, and down quark possess the ratio $1:2:3$ and hence they can be assigned the `square-root-mass numbers' $(1/3, 2/3, 1)$ respectively, these being in the inverse order as the ratios of their electric charge. The $e/\sqrt{m}$ ratios for the three particles then have the respective values $(3, 1, 1/3)$, whereas $e\sqrt{m}$ has the respective values $(1/3, 4/9, 1/3)$. The choice of square-root of mass as being more fundamental than mass is justified by recalling that in our approach, gravitation is derived from `squaring' an underlying spin one Lorentz interaction \cite{Singh2020DA}. It is reasonable then to assume that the spin one Lorentz interaction is sourced by $\sqrt{m}$, and to try to understand the origin of the square-root of the mass ratios, rather than origin of  the mass ratios themselves.

The above proposed quantised root-mass-ratios for the first generation has been justified in \cite{Vatsalya1} from a L-R symmetric extension of the standard model. [We also justify this aspect in detail in  \cite{vvs}, where we consider an $SU(3)$ gravi-color symmetry for gravitation, analogous to $SU(3)_{color}$ for QCD, and actually demonstrate a square-root mass ratio 1:2:3 for electron, up quark and down quark.]  A justification might come from the following. The automorphism group $G_2$ of the octonions has the two maximal subgroups $SU(3)$ and $SO(4)$. These two groups have an intersection $U(2)\sim SU(2)\times U(1)$. The $SU(3)$ is identified with $SU(3)_c$, the $SU(2)$ with the weak symmetry, and the $U(1)$ with $U(1)_{em}$. Thus the $U(1)_{em}$ is a subset also of the maximal sub-group $SO(4)$ which led us to propose the Lorentz-Weak-Electro symmetry, and hence this $U(1)$ might also determine the said quantised root-mass-ratios $(1/3, 2/3, 1)$ for the positron, up quark, and down quark respectively.  This implies, assuming a mass $0.511$ MeV for the electron, a consequent predicted mass of $2.044$ MeV for the up quark, and a predicted mass $4.599$ MeV for the down quark.

If we assume that the $e/\sqrt{m}$ ratios for the first generation of the charged fermions are absolute values [valid prior to the enormous scaling down of mass] then we can assign a root-mass number $e/3$ to the positron [and hence a mass number $e^2/9$], where the electric charge  $e$ is as given in Eqn. (\ref{echarge}). Hence the mass-number for the positron/electron is
\begin{equation}
\sqrt{G_{N}} \; m_{e+} = (1/1024) \exp [ 2/9 - 1/\sqrt{6} ] \ ( \hbar L_P / t_P )^{1/2}
\label{echarge}
\end{equation}
where $G_N$ is Newton's gravitational constant. Thus the mass number of the electron  is $1/(137\times 9)$ of Planck mass and has to be scaled down by the factor $f=2\times 10^{19}$ before it acquires the observed mass of $0.5$ MeV. This then is also the universal factor by which the assigned mass number of every quark and charged lepton must be scaled down to get it to its current value. This is not far from the twenty orders of mass-scale-down by the Karolyhazy effect in cosmology, proposed earlier in this section. The initial ratio of the electrostatic to gravitational attraction between an electron and a positron is $e^2 / (e^4/81) \sim 137\times 81 \sim 10^4$.

Since the square-root-mass ratio of the anti-down quark has been set to unity, and predicted above to be $4.599$ MeV (= $9 \times 0.511$ MeV), we will calculate the square-root-mass ratios of the other particles with respect to the anti-down-quark, and demonstrate a correlation of these ratios with the Jordan eigenvalues. Also, since a negative Jordan eigenvalue is to be associated with minus of square-root mass, for finding the mass-ratio, we take the absolute value of the anti-down-quark eigenvalue, which is negative. 

$\bullet$ Anti-muon : Take the ratio of the Jordan eigenvalues for the electron and the muon [see  Fig. 3]. Multiply by a factor representing the down quark (the first factor in the expression below). Then compare the resulting  value with the square-root mass ratio of the muon mass with respect to the electron mass:
\begin{equation}
    \frac{1+\sqrt{3/8}}{1 -\sqrt{3/8}}\; \times \; \frac{1/3+\sqrt{3/8}} {|1/3-\sqrt{3/8}|} =  14.10 \ ;\quad\ \sqrt{206.7682830} = 14.38
\end{equation}
$\bullet$ Anti-tau lepton : Using the eigenvalues for the charged leptons, we get the ratio for tau-lepton to electron:
\begin{equation}
    \frac{1+\sqrt{3/8}}{1 -\sqrt{3/8}} \; \times \; \frac{1/3+\sqrt{3/8}}{|1/3-\sqrt{3/8}|}\; \times  \;    \frac{1+\sqrt{3/8}}{1 -\sqrt{3/8}}  = 58.64\ ; \ \sqrt\frac{1776.86}{.511} = 58.97
\end{equation} 
 
$\bullet$ Charm quark with respect to up quark: This ratio is same as  the ratio of charm / up in Eqn. (\ref{charmr}).
\begin{equation}
 \frac{2/3 + \sqrt{3/8}} {2/3-\sqrt{3/8}} \ =  23.56 \ ;\quad\ \sqrt{\frac{1275}{2.3}} = 23.55
\end{equation}

$\bullet$ Top quark with respect to up quark: Again this ratio is analogous to the one for top / up in Eqn. (\ref{topr}).
\begin{equation}
    \frac{ {2/3} + \sqrt{3/8} } {  {2/3} - \sqrt{3/8} }  \; \times\;  \frac{ {2/3}  } {  {2/3} - {\sqrt{3/8} } }  = 289.26 \ ;\quad\ \sqrt{\frac{173210}{2.3}} = 274.42
\end{equation}

$\bullet$ Anti-strange quark with respect to down quark:
\begin{equation}
    \frac{1+\sqrt{3/8}}{1 -\sqrt{3/8}} \times 1 = 4.16   \ ; \quad\sqrt\frac{95}{4.7} = 4.50
\end{equation} 

$\bullet$ Anti-bottom quark with respect to down quark:
\begin{equation}
   \frac{1+\sqrt{3/8}}{1 -\sqrt{3/8}} \;  \times \; \frac{1+\sqrt{3/8}}{1}\; \times \;   \frac{1+\sqrt{3/8}}{1 -\sqrt{3/8}} = 27.90                   \ ;        \quad\sqrt\frac{4180}{4.7} = 29.82
\end{equation} 
These ratios made from the Jordan eigenvalues suggest a possible correlation with the square-root mass ratios, and hence provide a plausible definition of a mass quantum number for standard model fermions. This definition is justified from trace dynamics and its Lagrangian, and is a property both of the octonionic algebra and of the Lagrangian. This is completely analogous to the fact that in the octonionic approach to the standard model, quantisation of electric charge is deduced from eigenvalues of the $U(1)_{em}$ operator made from the Clifford algebra $Cl(6)$. Hence, square-root of mass is treated on the same footing as electric charge: their quantisation is a property of the algebra, not of the dynamics. The difference between charge quantisation and mass quantisation is that for finding the mass eigenstates, all three generations must be considered together, not one at a time. For a more detailed justification of this derivation of mass-ratios, including the motivation from the group $E_6$ and the connection with the definition of mass as Casimir of Poincar\'e algebra, the reader is refereed to the detailed review by Singh \cite{Singhreview}. 

The square-root mass numbers for the charged fermions are shown in Fig. 8. These have the same fundamental status as quantised electric charge values 1/3, 2/3 and 1.
\begin{figure}[!htb]
        \center{\includegraphics[width=\textwidth]
        {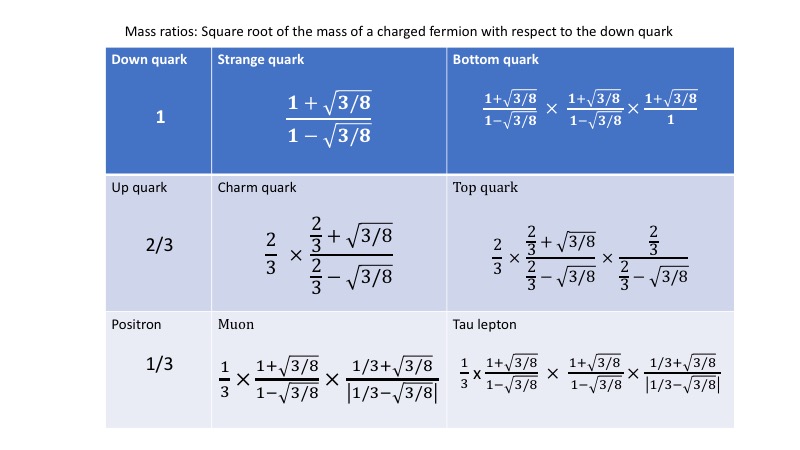}}
        \caption{The square-root mass numbers for charged fermions. These have the same fundamental status as quantised electric charge values 1/3, 2/3 and 1.}
 \end{figure}

\bigskip

\bigskip

\subsection { Quantum non-locality} Additional internal spatial dimensions which are not compact, yet very thin, offer a promising resolution to the quantum non-locality puzzle, thereby lifting the tension with 4D special relativity. Let us consider once again Baez's cube of Fig. 4. Any of the three quaternionic spaces containing the unit element 1 can play the role of the emergent 4D classical space-time
in which classical systems evolve. Let us say this classical universe is the plane $(1 e_6 e_1 e_5)$. Now, the true universe is the full 8D octonionic universe, with the four internal dimensions being probed [only by] quantum systems. Now we must recall that these four internal dimensions are extremely thin, of the order of Fermi dimensions, and along these directions no point is too far from each other, even if their separation in the classical 4D quaternion plane is billions of light years! Consider then, that Alice at 1 and Bob at $e_1$ are doing space-like separated measurements on a quantum correlated pair. Whereas the event at $e_1$ is outside the light cone of $1$, the correlated pair is always within each other's quantum wavelength along the internal directions, say the path $(1 e_3 e_2 e_7 e_1)$. The pair influences each other along this path acausally, because this route is outside the domain of 4D Lorentzian spacetime and its causal light-cone structure. The internal route is classically forbidden but allowed in quantum mechanics. This way neither special relativity nor quantum mechanics needs to be modified. It is also interesting to ask if evolution in Connes time in this 8D octonionic universe obeying generalised trace dynamics can violate the Tsirelson bound.

The exceptional Jordan algebra is of significance also in superstring theory, where it has been suggested that there is a relation between the EJA and the vertex operators of superstrings, and that the vertex operators represent couplings of strings \cite{corrigan, goodard}. This intriguing connection between the EJA, string theory and aikyon theory deserves to be explored further.

Lastly we mention that the Lagrangian (\ref{lagba}) that we have been studying closely resembles the Bateman oscillator \cite{Bateman} model, for which the Lagrangian is
\begin{equation}
L = m\dot{x} \dot{y} + \gamma (x \dot{y} - \dot{x} y) - k xy
\end{equation}
I thank Partha Nandi for bringing this fact to my attention. Considering that the Bateman oscillator represents a double oscillator with relative opposite signs of energy for the two oscillators undergoing damping, it is important to understand the implications for our theory. In particular, could this imply a cancellation of zero point energies between bosonic and fermionic modes, thus annulling the cosmological constant? And also whether this damping is playing any possible role in generating matter-anti-matter asymmetry?

\vskip 0.4 in

\noindent{\bf Acknowledgements}: I would like to thank Carlos Perelman for discussions and helpful correspondence, and for making me aware of the beautiful work of Dray and Manogue on the Jordan eigenvalue problem. I also thank Vivan Bhatt, Tanmoy Bhattacharya,  Cohl Furey, Niels Gresnigt, Garrett Lisi, Nehal Mittal, Rajrupa Mondal, Roberto Onoforio, Robert Wilson and Vatsalya Vaibhav for useful correspondence and discussions.

\vskip 0.2 in

\section{APPENDIX: Physical motivation for the present  theory: quantum (field) theory without classical time, as a route to quantum gravity and unification}
\noindent In this appendix, we recall from earlier work \cite{Singh2020DA} the motivation for developing a formulation of quantum theory without classical time, and how doing so leads to a pre-quantum, pre-spacetime theory which is a candidate for unification of general relativity with the standard model.
\subsection{Why there must exist a formulation of quantum theory which does not refer to classical time?
And
why such a formulation must exist at all energy scales, not just at the Planck energy scale.}
Classical time, on which quantum systems depend for a description of their evolution, is part of a classical space-time. Such a space-time - the manifold as well as the metric that overlies it - is produced by macroscopic bodies. These macroscopic bodies are a limiting case of quantum systems. In principle one can imagine a universe in which there are no macroscopic bodies, but only microscopic quantum systems. And this need not be just at the Planck energy scale.

As a thought experiment, consider an electron in a double slit interference experiment, having crossed the slits, and not yet reached the screen. It is in a superposed state, as if it has passed through both the slits. We want to know, 
non-perturbatively,  what is the spacetime geometry produced by the electron? Furthermore, we imagine that every macroscopic object in the universe is suddenly separated into its quantum, microscopic, elementary particle units. We have hence lost classical space-time! Perturbative quantum gravity is no longer possible. And yet we must be able to describe what gravitational effect the electron in the superposed state is producing. This is the sought for quantum theory without classical time! And the quantum system is at low non-Planckian energies, and is even non-relativistic. 
This is the sought for formulation we have developed, assuming only three fundamental constants a priori: Planck length $L_P$, Planck time $t_P$, and Planck's constant $\hbar$. Every other dimensionful constant, e.g. electric charge, and particle masses, are expressed in terms of these three. This new theory is a pre-quantum, pre-spacetime theory, needed even at low energies.

A system will be said to be a Planck scale system if any dimensionful quantity describing the system and made from these three constants, is order unity. Thus if time scales of interest to the system are order $t_{P} = 10^{-43}$ s, the system is Planckian. If length scales of interest are order $L_P = 10^{-33}$ cm, the system is Planckian. If speeds of interest are of the order $L_P/t_P = c = 3\times10^8$ cm/s then the system is Planckian. If the energy of the system is of the order $\hbar / t_P = 10^{19}$ GeV, the system is Planckian. If the action of the system is of the order $\hbar$, the system is Planckian. If the charge-squared  is of the order $\hbar c$, the system is Planckian. Thus in our concepts, the value 1/137 for the fine structure constant, being order unity in the units  $\hbar c$, is Planckian. This explains why this pre-quantum, pre-spacetime theory knows the low energy fine structure constant.

A quantum system on a classical space-time background is hugely non-Planckian. Because the classical space-time is being produced by macroscopic bodies each of which has an action much larger than $\hbar$. The quantum system treated in isolation is Planckian, but that is strictly speaking a very approximate description. The spacetime background cannot be ignored - only when the background is removed from the description, the system is exactly Planckian. This is the pre-quantum, pre-spacetime theory.

It is generally assumed that the development of quantum mechanics, started by Planck in 1900, was completed in the 1920s, followed by generalisation to relativistic  quantum field theory. This assumption, that the development of quantum mechanics is complete, is not necessarily correct - quantisation is not complete until the last of the classical elements - this being classical space-time - has been removed from its formulation.

The pre-quantum, pre-spacetime theory achieves that, giving also an anticipated theory of quantum gravity. What was not anticipated was that removing classical space-time from quantum theory will also lead to unification of gravity with the standard model. And yield an understanding of where the standard model parameters come from. It is clear that the sought for theory is not just a high energy Beyond Standard Model theory. It is needed even at currently accessible energies, so at to give a truly quantum formulation of quantum field theory. Namely, remove classical  time from quantum theory, irrespective of the energy scale. Surprisingly, in doing so, we gain answers to unsolved low energy aspects of the standard model and of gravitation.

The process of quantisation works very successfully for non-gravitational interactions, because they are not concerned with space-time geometry. However, it is not  necessarily correct to apply this quantisation process to spacetime geometry. Because the rules of quantum theory have been written by assuming a priori that classical time exists. How then can we apply these quantisation rules to classical time itself? Doing so leads to the notorious problem of time in quantum gravity - time is lost, understandably.
We do not quantise gravity. We remove classical space-time / gravity from quantum [field] theory. Space-time and gravity emerge as approximations from the pre-theory, concurrent with the emergence of classical macroscopic bodies. In this emergent universe, those systems which have not become macroscopic, are described by the beloved quantum theory we know - namely quantum theory on a classical spacetime background. This is an approximation to the pre-theory: in this approximation, the contribution of the said quantum system to the background spacetime is [justifiably] neglected.

\subsection{Why a quantum theory of gravity is needed at all energy scales, and not just at the Planck energy scale? And how that leads us to partially redefine what is meant by Planck scale: Replace Energy by Action.}
We have argued above that there must exist a formulation of quantum theory which does not refer to classical time. Such a formulation must in principle exist at all energy scales, not just at the Planck energy scale. For instance, in today's universe, if all classical objects were to be separated out into elementary particles, there would be no classical space-time and we would need such a formulation. Even though the universe today is a low energy universe, not a Planck energy universe.

Such a formulation is inevitably also a quantum theory of gravity. Arrived at, not by quantising gravity, but by removing classical gravity from quantum theory. We can also call such a formulation pure quantum theory, in which there are no classical elements: classical space-time has been removed from quantum theory. We also call it a pre-quantum, pre-spacetime theory.

What is meant by Planck scale, in this pre-theory?

Conventionally, a phenomenon is called Planck scale if: the time scale $T$ of interest is of the order Planck time $t_P$; and/or length scale $L$  of interest is of the order of Planck length  $L_P$; and/or energy scale $E$ of interest  is of the order Planck energy $E_P$. According to this definition of Planck scale, a Planck scale phenomenon is quantum gravitational in nature.
Since the pre-theory is quantum gravitational, but not necessarily at the Planck energy scale, we must partially revise the above criterion, when going to the pre-theory: replace the criterion on energy $E$ by a criterion on something else. This something else being the action of the system!

In the pre-theory, a phenomenon is called Planck scale if: the time scale $T$ of interest is of the order Planck time $T_P$; and/or length scale $L$  of interest is of the order of Planck length $L_P$; and/or the action $S$ of interest  is of the order Planck constant $\hbar$. According to this definition of Planck scale, a Planck scale phenomenon is quantum gravitational in nature.

Why does this latter criterion make sense? If every degree of freedom has an associated action  of order $\hbar$, together the many degrees of freedom cannot give rise to a classical spacetime. Hence,  even if the time scale T of interest and length scale L of interest are NOT Planck scale,  the system is quantum gravitational in nature. The associated energy scale $\hbar / T$ for each degree of freedom is much smaller than Planck scale energy $E_P$.  Hence in the pre-theory the criterion for a system to be quantum gravitational is DIFFERENT from conventional approaches to quantum gravity. And this makes all the difference to the formulation and interpretation of the theory. e.g. the low energy fine structure constant 1/137 is a Planck scale phenomenon [according to the new definition] because the square of the electric charge is order unity in the units $\hbar c = \hbar L_P / t_P$

In our pre-theory, there are three, and only three, fundamental constants: Planck length $L_P$, Planck time $t_P$ and Planck action $\hbar$. Every other parameter, such as electric charge, Newton's gravitational constant, standard model coupling constants, and masses of elementary particles, are defined and derived in terms of these three constants: $\hbar, L_P $ and $t_P$.  

In the pre-theory the universe is an 8D octonionic universe, as shown in the Fig.  3, the octonion, reproduced below in Fig. 9. The origin $e_0=1$ stands in for the real part of the octonion [coordinate time] and the other seven vertices stand in for the seven imaginary directions. A degree of freedom [i.e. `particle' or an atom of space-time-matter (STM)] is described by a matrix $q$ which resides on the octonionic space: $q$ has eight coordinate components $q_i$ where each $q_i$ is a matrix. We have replaced a four-vector in Minkowski space-time by an eight-matrix in octonionic space: and this describes the particle / STM atom. The STM atom evolves in Connes time, this time being over and above the eight octonionic coordinates. Its action is that of a free particle in this space: time integral of kinetic energy, the latter  being the square of velocity $\dot q$, where dot is derivative with respect to Connes time.  Eight octonionic coordinates are equivalent to ten Minkowski coordinates, because of $SL(2,O) \sim {\rm Spin}(9,1)$. 
\begin{figure}[!htb]
        \center{\includegraphics[width=\textwidth]
        {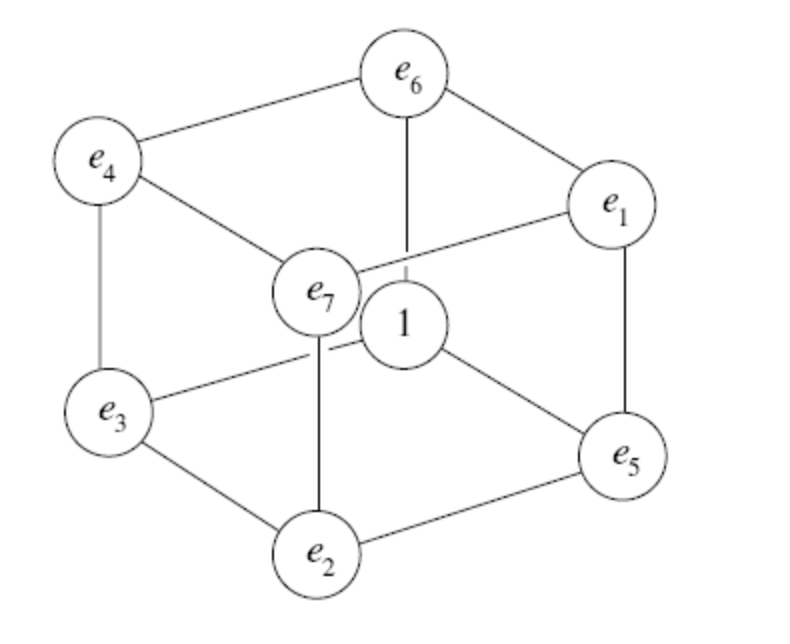}}
        \caption{\label{fig:my-label} The octonions [From Baez \cite{baez2001octonions}].}
      \end{figure}
The symmetries of this space  are the symmetries of the (complexified) octonionic algebra: they contain within them the symmetries of the standard model, including the 4D-Lorentz symmetry.

The classical 4D Minkowski universe is one of the three planes (quaternions) intersecting at the origin $e_0 = 1$. Incidentally the three lines originating from $e_0$ represent complex numbers. The four imaginary directions not connected to the origin represent directions along which the standard model forces lie (internal symmetries). Classical systems live on the 4D quaternionic plane. Quantum systems (irrespective of whether they are at Planck energy scale) live on the entire 8D octonion. Their dynamics is the sought for quantum theory without classical time. This dynamics is oblivious to what is happening on the 4D classical plane. QFT as we know it is this pre-theory projected to the 4D Minkowski space-time. The present universe has arisen as a result of a symmetry breaking in the 8D octonionic universe: the electroweak symmetry breaking. Which in this theory is actually the color-electro -- weak-Lorentz symmetry breaking. Classical systems condense on to the 4D Minkowski plane as a result of spontaneous localisation, which precipitates the electro-weak symmetry breaking in the first place. The fact that weak is part of weak-lorentz should help understand why the weak interaction violates parity, whereas electro-color does not. Hopefully the theory will shed some light also on the strong-CP problem.

\subsection{What is Trace Dynamics? : Trace dynamics is quantisation, without imposing  the Heisenberg algebra}

In the conventional development of canonical quantisation, the two essential steps are:

1. Quantisation Step 1 is to raise classical degrees of freedom, the real numbers  $q$ and $p$, to the status of operators / matrices.  This is a very reasonable thing to do.

2. Quantisation Step 2 is very restrictive! Impose the Heisenberg algebra $[q, p] = i \hbar$.  Its only justification  is that the theory it gives rise to is extremely successful and consistent with every experiment done to date. 
In classical dynamics, the initial values of $q$ and $p$ are independently prescribed. There is NO relation between the initial $q$ and $p$. Once prescribed initially, their evolution is determined by the dynamics. Whereas, in quantum mechanics, a theory supposedly more general than classical mechanics, the initial values of the operators $q$ and $p$ must also obey the constraint  $[q, p] = i \hbar$. This is highly restrictive!

3. It would be more reasonable if there were to be a dynamics based only on Quantisation Step 1. And then Step 2 emerges from this underlying dynamics in some approximation. This is precisely what Trace Dynamics is. Only step 1 is applied to classical mechanics. $q$ and $p$ are matrices, and the Lagrangian is the trace of a matrix polynomial made from $q$ and its velocity. The matrix valued equations of motion follow from variation of the  trace Lagrangian. They describe dynamics. This is the theory of trace dynamics developed by Adler \cite{Adler:94, AdlerMillard:1996, Adler:04} - a pre-quantum theory, which we have generalised  to a 
pre-quantum, pre-spacetime theory \cite{maithresh2019}. 

4. This matrix valued dynamics, i.e. trace dynamics, is more general than quantum field theory, and assumed to hold at the Planck scale, and also whenever background classical spacetime is absent, no matter what the energy scale.  The Heisenberg algebra is shown to emerge at lower energies, or when space-time emerges, after coarse-graining the trace dynamics over length scales much larger than Planck length scale. Thus, quantum theory is midway between trace dynamics and classical dynamics. 

5. The moral of the story is that we assume that  quantum field theory does not hold at the Planck scale. Trace dynamics does. QFT is emergent.

6. The other assumption one makes at the Planck scale is to replace the 4-D classical spacetime manifold by an 8D octonionic spacetime manifold, so as to obtain a canonical definition of spin. This in turn allows for a Kaluza-Klein type unification of gravity and the standard model. Also, an 8D octonionic spacetime is equivalent to a 10-D Minkowski space-time. It is very rewarding to work with 8D octonionic, rather than 10D Minkowski - the symmetries manifest much more easily.

7. Trace dynamics plus octonionic spacetime together give rise to a highly promising avenue for constructing a theory of quantum gravity, and of unification. 4D classical spacetime obeying GR emerges as an approximation at lower energies, alongside the emergent quantum theory.

8. How is this different from string theory? In many ways it IS like string theory, but ${\it without}$ the Heisenberg algebra! The gains coming from  dropping $[q,p]=i\hbar$ at the Planck scale are enormous. One now has a non-perturbative description of pre-space-time at the Planck scale.
The symmetry principle behind the unification is very beautiful: physical laws are invariant under algebra automorphisms of the octonions. This unifies the internal gauge transformations of the standard model with the 4D spacetime diffeomorphisms of general relativity. The automorphism group of the octonions, the Lie group $G_2$, which is the smallest of the five exceptional Lie groups, contains within itself the symmetries  $SU(3) \times SU(2) \times U(1)$ of the standard model, along with the Lorentz symmetry. The free parameters of the standard model are determined by the characteristic equation of the exceptional Jordan algebra $J_3(O)$, whose automorphism group $F_4$ is the exceptional Lie group after $G_2$.

\subsection{Normed division algebras, trace dynamics, and relativity in higher dimensions. And how these relate to quantum field theory and the standard model}
Let us consider the four normed division algebras ${\mathbb R, \mathbb C, \mathbb H, \mathbb O}$  [Reals, Complex Numbers, Quaternions, Octonions] in the context of the space-times associated with them, and how these algebras relate to trace dynamics. This can be understood graphically with the help of Fig. 7 above, which contains within itself a representation of all the four division algebras. 

Let us start with the reals $\mathbb R$, represented in the above diagram by the origin $e_0 = 1$. This direction represents the time coordinate, in all the four different space-times associated with these four division algebras. The three lines emanating from the origin and connecting respectively to $e_3, e_5, e_6$ represent complex numbers $\mathbb C$. The three planes intersecting at the origin represent quaternions $\mathbb H$ and the full cube represents the octonions $\mathbb O$.

\bigskip

\noindent {\it Galilean relativity and Newtonian mechanics:} This is related to the quaternions, and we assume each of the three planes intersecting at the origin represent absolute Newtonian space [say in the plane ($1, e_1, e_5, e_6$) we set $e_1 = \hat x, e_5 = \hat y, e_6 = \hat z$]. Galilean invariance is assumed, and the spatial symmetry group is $SO(3)$, the group of rotations in three dimensional space; this is also the automorphism group $Aut(\mathbb H)$ of the quaternions. The origin represents absolute Newtonian time, and we have Newtonian dynamics in which the action principle for the free particle represented by the configuration variable ${\bf q}$, which is a three-vector,  is simply
\begin{equation}
S = \int dt \; \dot{\bf q}^2
\end{equation}
The generalisation to many-particle systems interacting via potentials is obvious and well-known. Newtonian gravity can be consistently described in this framework. The dynamical variables, being real-number valued three-vectors, all commute with each other. The important approximation made in the physical space is that by hand we set $e_1^2 = e_2^2 = e_3^2 = 1$, instead of $-1$. This of course is what gives us the Newtonian absolute space (Euclidean geometry) and absolute time, and the manifold $R^3$ for physical space. The associated algebra is $\mathbb R \times \mathbb H$, in an approximate sense, which becomes precise only in special relativity, as discussed below.

[The algebra $\mathbb C$ represents a 2D physical space, and $\mathbb R \times \mathbb C$ represents a space-time for Newtonian mechanics in absolute two-space represented by $\mathbb C$, and absolute time $\mathbb R$. The homomorphism $SL(2, \mathbb R)\sim SO(2,1)$ suggests that we can relate 2x2 real-valued matrices to a 2+1 relativistic space-time. This observation becomes very relevant when we relate normed division algebras to relativity.]

To go from here to trace dynamics, we will raise all dynamical variables from three-vectors to three-matrices. Thus ${\bf \hat q}$ is a matrix-valued three-vector whose three spatial components ${\bf \hat q_1, \hat q_2, \hat q_3}$ are matrices whose entries are real numbers. The Lagrangian for a free particle will now be the trace of the matrix polynomial 
${\bf \dot  {\hat q}}^2$, and hence the action is
\begin{equation}
S = \int dt \; Tr [{\bf \dot  {\hat q}}^2]
\end{equation}
The underlying three-space continues to have the symmetry group $SO(3)$ and the dynamics obeys Galilean invariance; this is implemented on the trace dynamics action via the unitary transformations generated by the generators of $SO(3)$.

\bigskip

\noindent {\it Special relativity, Complex quaternions, and the algebra $\mathbb R \times  \mathbb C \times \mathbb H$}:

Consider the quaternionic four vector ${\bf x} = x_0 e_0 + x_1 e_1 + x_2 e_2 + x_4 e_4$ and the corresponding position four-vector for a particle in special relativity: ${\bf q_i} = q_0 e_0 + q_1 e_1 + q_2 e_2 + q_4 e_4$. One can define the four-metric on this Minkowski space-time whose symmetry group is the Lorentz group $SO(3,1)$ having the universal cover Spin(3,1) isomorphic to $SL(2,C)$. The complex quaternions generate the boosts and rotations of the Lorentz group SO(3,1). They can be used to obtain a faithful representation of  the Clifford algebra $Cl(2)$ and fermionic ladder operators constructed from this algebra can be used to generate the Lorentz algebra $SL(2,\mathbb C)$. Also, $Cl(2)$ can be used to construct left and right handed Weyl spinors as minimal left ideals of this Clifford algebra, and as is well known the Dirac spinor and the Majorana spinor can be defined from the Weyl spinors. Cl(2) also gives the vector and scalar representations of the Lorentz algebra. These results are lucidly described in Furey's Ph. D. thesis \cite{f1, f2, f3} as well as also in her video lecture series on standard model and division algebras https://www.youtube.com/watch?v=GJCKCss43WI\&ab_channel=CohlFureyCohlFurey

The above relation between the Clifford algebra $Cl(2)$ and the Lorentz algebra $SL(2,C)$ strongly suggests, keeping in view the earlier conclusions for $Cl(6)$ and the standard model and the octonions \cite{f1, f2, f3}, that the $Cl(2)$ algebra describes the left handed neutrino and the right-handed anti-neutrino, and a pair of spin one Lorentz bosons. This is confirmed by writing the following trace dynamics Lagrangian and action on the quaternionic space-time of special relativity, thereby generalising the relativistic particle $S = - mc \int ds$:
\begin{equation}
\frac{S}{C_0} =  \frac{a_0}{2} \int \frac{d\tau}{\tau_{Pl}} \; Tr  \bigg[\dot{q}_B^{\dagger} + i\frac{\alpha}{L} q_B^\dagger+ a_0 \beta_1\left( \dot{q}_F^\dagger  + i\frac{\alpha}{L} q_F^\dagger\right)\bigg] \times \bigg[ \dot{q}_B + i\frac{\alpha}{L} q_B+ a_0 \beta_2\left( \dot{q}_F + i\frac{\alpha}{L} q_F\right)\bigg] 
\label{ymi}
\end{equation} 
where  $a_0 \equiv L_P^2 / L^2$. 
This Lagrangian is identical in form to the one studied earlier in the present paper, but with a crucial difference that it is now written on 4D quaternionic space-time, not on 8D octonionic space-time. Thus $\dot{q}_B$ and $q_B$ have four components between them, not eight: $q_B = q_{Be2}\; e_2 + q_{Be4} \; e_4; \;  \dot{q}_B = \dot{q}_{Be0} \; e_0 + \dot{q}_{Be1}\ e_1$. Similarly, the fermionic matrices have four components between them, not eight. Thus $q_F = q_{Fe2}\; e_2 + q_{Fe4} \; e_4; \;  \dot{q}_F = \dot{q}_{Fe0} \; e_0 + \dot{q}_{Fe1}\ e_1$

This has far-reaching consequences. Consider first the case where we set $\alpha=0$. The Lagrangian then is
\begin{equation}
\frac{S}{C_0} =  \frac{a_0}{2} \int \frac{d\tau}{\tau_{Pl}} \; Tr  \bigg[\dot{q}_B^{\dagger} + a_0 \beta_1 \dot{q}_F^\dagger  \bigg] \times \bigg[ \dot{q}_B + a_0 \beta_2 \dot{q}_F \bigg] 
\label{ymi2}
\end{equation} 
By opening up the terms into their coordinate components, the various degrees of freedom can be identified with the Higgs, the Lorentz bosons, the neutral weak isospin boson, and two neutrinos. The associated space-time symmetry is the Lorentz group $SO(3,1)$ and the associated Clifford algebra is $Cl(2)$, reminding us again of the homomorphism $SL(2,\mathbb C)\sim SO(3,1)$.

When $\alpha$ is retained, the Lagrangian describes Lorentz-weak symmetry of the leptons: electron, positron, two neutrinos of the first generation, the Higgs, two Lorentz bosons, and the three weak isospin bosons. To our understanding, the associated Clifford algebra is still $Cl(2)$ but now all the quaternionic degrees of freedom have been used in the Lagrangian and in the construction of the particle states.. What we likely have here is the extension of the Lorentz algebra by an $SU(2)$, as shown in Figure 10 below, borrowed from our earlier work \cite{Singh2020DA}. It remains to be understood if now the homomorphism $SL(2, \mathbb H)\sim SO(5,1)$ comes into play. And also, whether a quaternionic triality \cite{FH} could explain the existence of three generations of leptons. These aspects are currently under investigation.
\begin{figure}[!htb]
        \center{\includegraphics[width=\textwidth]
        {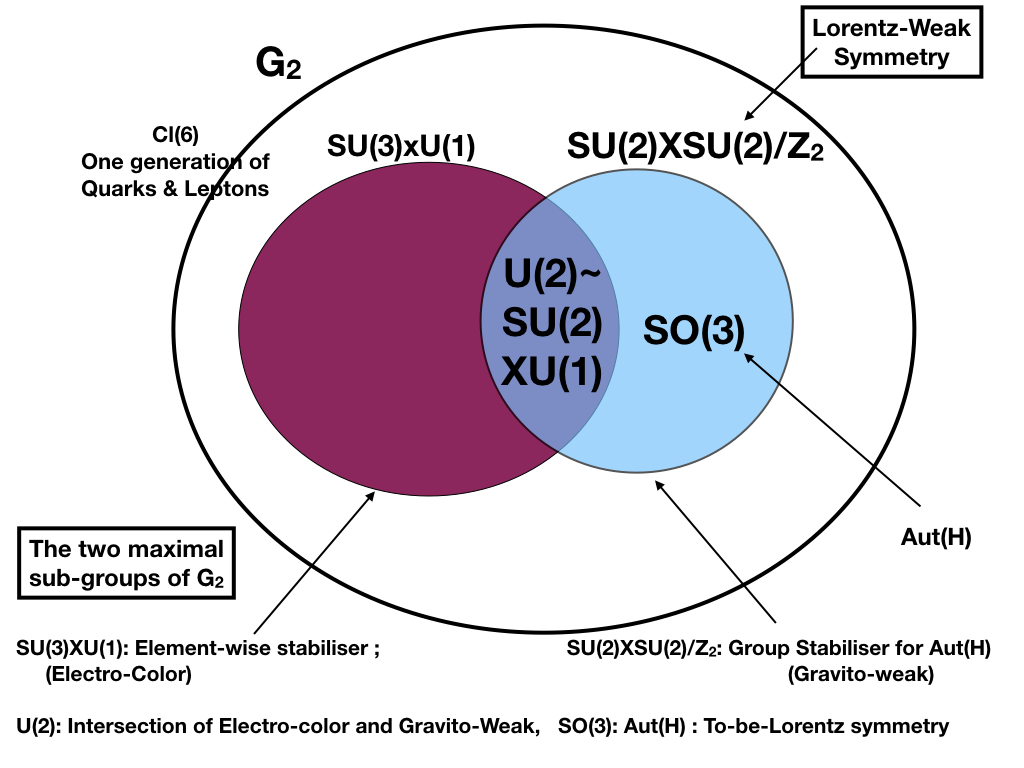}}
        \caption{\label{fig:my-label} The maximal sub-groups of $G_2$ and their intersection [From Singh \cite{Singh2020DA}]. Strictly speaking, the maximal subgroups of $G_2$ are $SU(3)$ and $SO(4)$. The $U(1)$ arises from the number operator made from generators of the Clifford algebra $Cl(6)$.}
      \end{figure}
      It is now only natural that this trace dynamics be extended to the last of the division algebras, the octonions, so as to construct an octonionic special relativity. This amounts to extending the Lorentz algebra by $U(3)$, as can be inferred from Fig. 10. 

\bigskip

\noindent{\it Octonionic special relativity, complex octonions, and the algebra $\mathbb R \times \mathbb C \times \mathbb H  \times \mathbb O$}

The background space-time is now an octonionic space-time with coordinate vector ${\bf x} = x_0 e_0 + x_1 e_1 + x_2 e_2 + x_4 e_4  + x_3 e_3 + x_5 e_5 + x_6 e_6 + x_7 e_7$, and the corresponding eight-vector for  a particle in this octonionic special relativity is ${\bf q_i} = q_0 e_0 + q_1 e_1 + q_2 e_2 + q_4 e_4 + q_3 e_3 + q_5 e_5 + q_6 e_6 + q_7 e_7$. In ordinary relativity, the $q_i$ are real numbers, but now in trace dynamics they are bosonic or fermionic matrices. The space-time symmetry group is the automorphism group $G_2$ of the octonions, shown in Fig. 10, along with its maximal sub-groups, which reveal the standard model along with its 4D Lorentz symmetry. The Lagrangian is the same as in (\ref{ymi2}) above, but now written on the 8D octonionic space-time. As a result, $q_B$ and $q_F$ have component indices (3, 5, 6, 7) whereas their time derivatives have indices (0, 1, 2, 4). This is the Lagrangian analysed in the main part of the present paper and it now includes quarks as well as leptons, along with all twelve standard model gauge bosons plus two Lorentz bosons.

We note the peculiarity that the weak part of the  Lorentz-weak symmetry of the leptons, obtained by extending the Lorentz symmetry, intersects with the electr-color sector provided by $U(3) \sim SU(3) \times U(1)$. This strongly suggests that the lepton part of the weak sector can be deduced from the electro-color symmetry. This is confirmed by the earlier work of Stoica \cite{Stoica}, Furey \cite{f3} and our own earlier work \cite{Singh2020DA}. 

We see that this Lagrangian is a natural generalisation of Newtonian mechanics and 4D special relativity to the last of the division algebras, the octonions, which represent a 10D Minkowski space-time because of the homomorphism $SL(2, \mathbb O) = SO(9,1)$.       

Left-Right symmetry breaking separates LH electric charge eigenstates from RH square-root mass particle eigenstates. The square-root of mass carries two signs: plus for matter, and minus for antimatter. CPT operations are mathematically defined as follows: complex conjugation for C, octonionic conjugation for P, and time-reversal operator T corresponds to change of sign of square-root mass. Pre-gravitation, i.e. $SU(2)_R$ symmetry, is a vector interaction which is attractive for matter-matter, attractive for anti-matter anti-matter, but repulsive for matter-antimatter. It is possible that the L-R symmetry breaking separates matter from antimatter in the very early universe, and this could be a possible explanation for the origin of matter-antimatter asymmetry. At the epoch of creation, our universe, made of matter, moves forward in time, whereas the antimatter universe moves backward in time. The two universes together obey CPT symmetry [for an earlier elegant proposal of CPT symmetric cosmology see \cite{BoyleTurok}). Prior to this symmetry breaking the universe is scale invariant.

\bigskip

\noindent{\it Emergent quantum field theory} 

In the entire discussion above, relating generalised trace dynamics to  the standard model, we have made no reference to quantum field theory. The pre-quantum, pre-space-time matrix-valued Lagrangian dynamics      which we have constructed above, reveals the standard model and its symmetries (including the Lorentz symmetry) without any fine tuning. Quantum field theory, and classical space-time, are emergent    from this pre-theory, after coarse-graining the underlying theory over time-scales much larger than Planck time, in the spirit of Adler's trace dynamics.

String theory is pre-space-time, but not pre-quantum.
Trace dynamics is pre-quantum, but not pre-space-time.
The octonionic theory [O-theory] is pre-space-time and pre-quantum. It generalises trace dynamics to a pre-quantum, pre-space-time theory.
The O-theory is not intended as an alternative to quantum field theory. Rather, it is applicable in those circumstances when a background classical time is not available for writing down the rules of QFT. Then, the O-theory also reveals itself to be pre-quantum.
When a background classical time becomes available, O-theory coincides with QFT and is no longer pre-quantum. 
O-theory reveals the symmetries of the standard model without any fine-tuning, and also shows a route for determining the free parameters of the standard model. This comes about because the background non-commutative spacetime fixes the properties of the allowed elementary particles. In this way, O-theory has a promising potential to tell us, in a mathematically precise way,  where the standard model, and classical space-time, come from.
The O-theory is not a Grand Unified Theory [GUTs]. GUTs determine internal symmetries by making specific choices for the internal symmetry group, while classical space-time and QFT are kept intact. In contrast to this, the O-theory retains neither QFT nor a classical space-time. The symmetries of O-theory are a unification of internal and spacetime symmetries, in the spirit of a Kaluza-Klein theory.

The diagram below in Fig. 11 lists the three main steps in which the octonionic theory is developed. Current investigation is focused at the third step.
\begin{figure}[!htb]
        \center{\includegraphics[width=\textwidth]
        {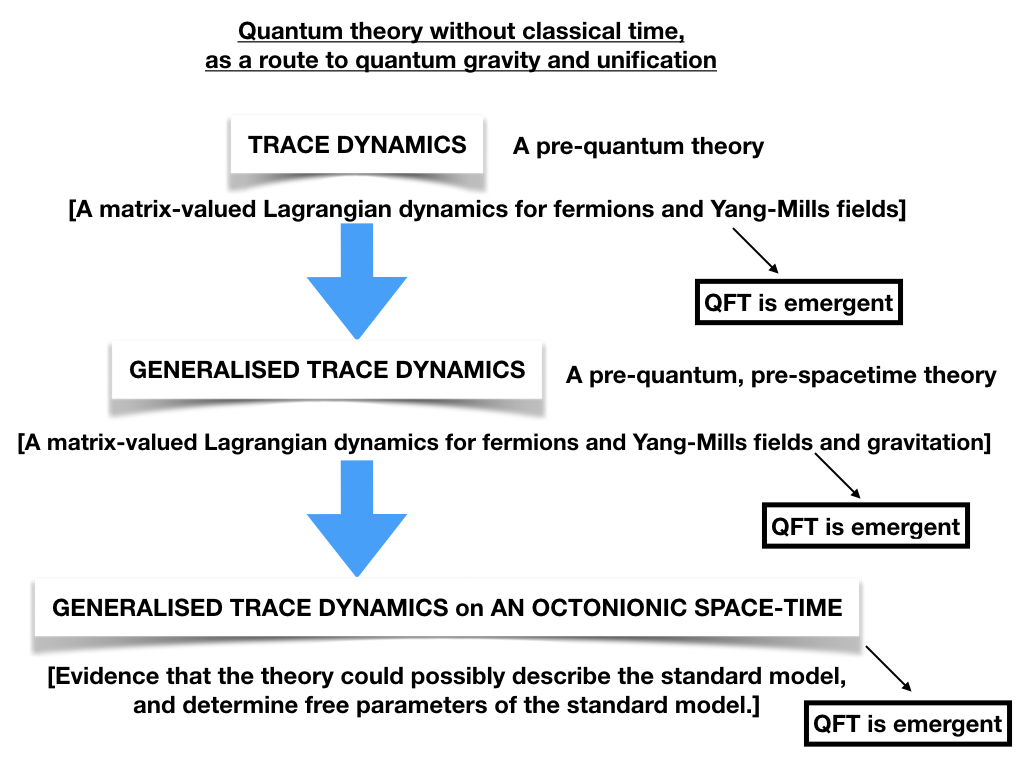}}
        \caption{\label{fig:my-label} The pre-space-time, pre-quantum octonionic theory in three key steps. The degrees of freedom are `atoms of space-time-matter' [STM]. An STM atom is an elementary fermion along with all the fields that it produces. The action for an STM atom resembles a 2-brane in a 10+1 dimensional Minkowski spacetime. The fundamental universe is made of enormously many STM atoms. From here, quantum field theory is emergent upon coarse-graining the underlying fundamental theory.}
      \end{figure}

The emergence of standard quantum field theory on a classical space-time background is a result of coarse-graining and spontaneous localisation and  has been described in our earlier papers \cite{maithresh2019, MPSingh}. Spontaneous localisation gives rise to macroscopic classical bodies and 4D classical space-time. From the vantage point of this space-time those STM atoms which have not undergone spontaneous localisation appear, upon coarse-graining of their dynamics,  as they are conventionally described by quantum field theory on a 4D classical space-time. Operationally, the transition from the action of the pre-spacetime pre-quantum theory is straightforward to describe.   Suppose the relevant term in the action of the pre-theory is denoted as $\int d\tau\;  \bigg[ Tr [T_1] + Tr [T_2] + Tr [T_3] \bigg]$. Say for instance the three terms respectively describe the electromagnetic field, the action of a $W$ boson on an electron, and the action of a gluon on an up quark. Then, the corresponding action for conventional QFT will be recovered as:
\begin{equation}
    \int  d\tau\; \bigg[ Tr [T_1] + Tr [T_2] + Tr [T_3] \bigg]  \quad  \longrightarrow      \quad   \int d\tau \; \int d^4x \bigg[ [T_{1QFT}] +  [T_{2QFT}] +  [T_{3QFT}] \bigg ]
\end{equation} 
The trace has been replaced by the space-time volume integral, and each of the three terms have correspondingly been replaced by the conventional field theory actions for the three cases: conventional action for the electromagnetic field, for the $W$ boson acting on the electron, and for the gluon acting on the up quark. In this way, QFT is recovered from the pre-theory. 

However, by starting from the pre-theory, we can answer questions which the standard model cannot answer. We know now why the standard model has the symmetries it does, and why the dimensionless free parameters of the standard model take the values they do. These are fixed by the algebra of the octonions which defines the 8D octonionic space-time in the pre-theory. While this is work in progress, it provides a promising avenue for understanding the origin of the standard model and its unification with gravitation.

\section{AUTHOR DECLARATIONS}

\noindent The author has no conflict of interests to declare.

\noindent No data has been used in the preparation of this manuscript.

\noindent The author acknowledges the support of the Department of Atomic Energy, Government of India, under Project Identification No. RTI4002.

\vskip 0.4 in

\centerline{\bf REFERENCES}

\bibliography{biblioqmtstorsion.bib}

\end{document}